\documentclass{nature}

\usepackage[british]{babel}
\usepackage[utf8]{inputenc}
\usepackage{babelbib}
\usepackage{url}
\usepackage{graphicx}
\usepackage{subfig}
\usepackage{calc}
\usepackage{floatflt}
\usepackage{amssymb, amsmath, amsthm}
\usepackage{tabularx}
\usepackage{multirow}
\usepackage{array}
\usepackage{xcolor}
\usepackage{bm}
\usepackage{stmaryrd}
\usepackage{stackrel}
\usepackage{algpseudocode}
\usepackage{algorithm}
\usepackage{rotating}
\usepackage{mwe,tikz}\usepackage[percent]{overpic}
\usepackage{float}
\usepackage{listings}

\usepackage{numprint}
\usepackage{memhfixc}
\usepackage{makecell}
\usepackage[hidelinks]{hyperref}

\usepackage{siunitx}
\usepackage{booktabs}

\usepackage{textcomp}
\usepackage{lineno}

\usepackage{mathtools}

\title{Neural \'{E}tendue Expander for Ultra-Wide-Angle High-Fidelity Holographic Display}
\author{Ethan Tseng$^{1}$, Grace Kuo$^{2}$, Seung-Hwan Baek$^{1,3}$, Nathan Matsuda$^{2}$, Andrew Maimone$^{2}$, Florian Schiffers$^{2}$, Praneeth Chakravarthula$^{1}$, Qiang Fu$^{4}$, Wolfgang Heidrich$^{4}$, Douglas Lanman$^{2}$, and Felix Heide$^{1\dagger}$}

\begin{document}

\maketitle



\definecolor{brightray}{rgb}{0.8,0.8,0.8}
\definecolor{Gray}{rgb}{0.5,0.5,0.5}
\definecolor{darkblue}{rgb}{0,0,0.7}
\definecolor{orange}{rgb}{1,.5,0} 
\definecolor{red}{rgb}{1,0,0} 
\definecolor{blue}{rgb}{0,0,1} 
\definecolor{darkgreen}{rgb}{0,0.7,0} 
\definecolor{darkred}{rgb}{0.7,0,0} 

\newcommand{\heading}[1]{\noindent\textbf{#1}}
\newcommand{\note}[1]{{{\textcolor{orange}{#1}}}}
\newcommand{\todo}[1]{{\textcolor{red}{TODO: #1}}}
\newcommand{\changed}[1]{{\textcolor{blue}{#1}}}
\newcommand{\removed}[1]{{\textcolor{brightray}{{#1}}}}
\newcommand{\revision}[1]{{{#1}}}
\newcommand{\place}[1]{ \begin{itemize}\item\textcolor{darkblue}{#1}\end{itemize}}
\newcommand{\de}{\mathrm{d}}

\newcommand{\baek}[1]{{\textcolor{darkgreen}{[BAEK: #1]}}}
\newcommand{\ethan}[1]{{\textcolor{darkred}{[ETHAN: #1]}}}
\newcommand{\felix}[1]{{\textcolor{darkblue}{[FELIX: #1]}}}

\newcommand{\BEAS}{\begin{eqnarray*}}
\newcommand{\EEAS}{\end{eqnarray*}}
\newcommand{\BEA}{\begin{eqnarray}}
\newcommand{\EEA}{\end{eqnarray}}
\newcommand{\BEQ}{\begin{equation}}
\newcommand{\EEQ}{\end{equation}}
\newcommand{\BIT}{\begin{itemize}}
\newcommand{\EIT}{\end{itemize}}
\newcommand{\BNUM}{\begin{enumerate}}
\newcommand{\ENUM}{\end{enumerate}}

\newcommand{\BA}{\begin{array}}
\newcommand{\EA}{\end{array}}

\newcommand{\eg}{{\it e.g.}}
\newcommand{\ie}{{\it i.e.}}
\newcommand{\etc}{{\it etc.}}

\newcommand{\ones}{\mathbf 1}

\newcommand{\reals}{{\mbox{\bf R}}}
\newcommand{\integers}{{\mbox{\bf Z}}}
\newcommand{\eqbydef}{\mathrel{\stackrel{\Delta}{=}}}
\newcommand{\complex}{{\mbox{\bf C}}}
\newcommand{\symm}{{\mbox{\bf S}}}  

\newcommand{\Span}{\mbox{\textrm{span}}}
\newcommand{\Range}{\mbox{\textrm{range}}}
\newcommand{\nullspace}{{\mathcal N}}
\newcommand{\range}{{\mathcal R}}
\newcommand{\Nullspace}{\mbox{\textrm{nullspace}}}
\newcommand{\Rank}{\mathop{\bf Rank}}
\newcommand{\Tr}{\mathop{\bf Tr}}
\newcommand{\diag}{\mathop{\bf diag}}
\newcommand{\lambdamax}{{\lambda_{\rm max}}}
\newcommand{\lambdamin}{\lambda_{\rm min}}

\newcommand{\Expect}{\mathop{\bf E{}}}
\newcommand{\Prob}{\mathop{\bf Prob}}
\newcommand{\erf}{\mathop{\bf erf}}

\newcommand{\Co}{{\mathop {\bf Co}}}
\newcommand{\co}{{\mathop {\bf Co}}}
\newcommand{\dist}{\mathop{\bf dist{}}}
\newcommand{\Ltwo}{{\bf L}_2}
\newcommand{\QED}{~~\rule[-1pt]{8pt}{8pt}}\def\qed{\QED}
\newcommand{\approxleq}{\mathrel{\smash{\makebox[0pt][l]{\raisebox{-3.4pt}{\small$\sim$}}}{\raisebox{1.1pt}{$<$}}}}
\newcommand{\epi}{\mathop{\bf epi}}

\newcommand{\vol}{\mathop{\bf vol}}
\newcommand{\Vol}{\mathop{\bf vol}}
\newcommand{\Card}{\mathop{\bf card}}

\newcommand{\dom}{\mathop{\bf dom}}
\newcommand{\aff}{\mathop{\bf aff}}
\newcommand{\cl}{\mathop{\bf cl}}
\newcommand{\Angle}{\mathop{\bf angle}}
\newcommand{\intr}{\mathop{\bf int}}
\newcommand{\relint}{\mathop{\bf rel int}}
\newcommand{\bd}{\mathop{\bf bd}}
\newcommand{\vect}{\mathop{\bf vec}}
\newcommand{\dsp}{\displaystyle}
\newcommand{\foequal}{\simeq}
\newcommand{\VOL}{{\mbox{\bf vol}}}
\newcommand{\xopt}{x^{\rm opt}}

\newcommand{\Xb}{{\mbox{\bf X}}}
\newcommand{\xst}{x^\star}
\newcommand{\varphist}{\varphi^\star}
\newcommand{\lambdast}{\lambda^\star}
\newcommand{\Zst}{Z^\star}
\newcommand{\fstar}{f^\star}
\newcommand{\xstar}{x^\star}
\newcommand{\xc}{x^\star}
\newcommand{\lambdac}{\lambda^\star}
\newcommand{\lambdaopt}{\lambda^{\rm opt}}

\newcommand{\geqK}{\mathrel{\succeq_K}}
\newcommand{\gK}{\mathrel{\succ_K}}
\newcommand{\leqK}{\mathrel{\preceq_K}}
\newcommand{\lK}{\mathrel{\prec_K}}
\newcommand{\geqKst}{\mathrel{\succeq_{K^*}}}
\newcommand{\gKst}{\mathrel{\succ_{K^*}}}
\newcommand{\leqKst}{\mathrel{\preceq_{K^*}}}
\newcommand{\lKst}{\mathrel{\prec_{K^*}}}
\newcommand{\geqL}{\mathrel{\succeq_L}}
\newcommand{\gL}{\mathrel{\succ_L}}
\newcommand{\leqL}{\mathrel{\preceq_L}}
\newcommand{\lL}{\mathrel{\prec_L}}
\newcommand{\geqLst}{\mathrel{\succeq_{L^*}}}
\newcommand{\gLst}{\mathrel{\succ_{L^*}}}
\newcommand{\leqLst}{\mathrel{\preceq_{L^*}}}
\newcommand{\lLst}{\mathrel{\prec_{L^*}}}

\newtheorem{theorem}{Theorem}[section]
\newtheorem{corollary}{Corollary}[theorem]
\newtheorem{lemma}[theorem]{Lemma}
\newtheorem{proposition}[theorem]{Proposition}

\newenvironment{algdesc}%
{\begin{quote}}{\end{quote}}

\def\figbox#1{\framebox[\hsize]{\hfil\parbox{0.9\hsize}{#1}}}

\makeatletter
\long\def\@makecaption#1#2{
   \vskip 9pt
   \begin{small}
   \setbox\@tempboxa\hbox{{\bf #1:} #2}
   \ifdim \wd\@tempboxa > 5.5in
        \begin{center}
        \begin{minipage}[t]{5.5in}
        \addtolength{\baselineskip}{-0.95pt}
        {\bf #1:} #2 \par
        \addtolength{\baselineskip}{0.95pt}
        \end{minipage}
        \end{center}
   \else
    \hbox to\hsize{\hfil\box\@tempboxa\hfil}
   \fi
   \end{small}\par
}
\makeatother

\newcounter{oursection}
\newcommand{\oursection}[1]{
 \addtocounter{oursection}{1}
 \setcounter{equation}{0}
 \clearpage \begin{center} {\Huge\bfseries #1} \end{center}
 {\vspace*{0.15cm} \hrule height.3mm} \bigskip
 \addcontentsline{toc}{section}{#1}
}
\newcommand{\oursectionf}[1]{  
 \addtocounter{oursection}{1}
 \setcounter{equation}{0}
 \foilhead[-.5cm]{#1 \vspace*{0.8cm} \hrule height.3mm }
 \LogoOn
}
\newcommand{\oursectionfl}[1]{  
 \addtocounter{oursection}{1}
 \setcounter{equation}{0}
 \foilhead[-1.0cm]{#1}
 \LogoOn
}

\newcommand{\Mat}[1]    {{\ensuremath{\mathbf{\uppercase{#1}}}}} 
\newcommand{\Vect}[1]   {{\ensuremath{\mathbf{\lowercase{#1}}}}} 
\newcommand{\Vari}[1]   {{\ensuremath{\mathbf{\lowercase{#1}}}}} 
\newcommand{\Id}				{\mathbb{I}} 
\newcommand{\Diag}[1] 	{\operatorname{diag}\left({ #1 }\right)} 
\newcommand{\Opt}[1] 	  {{#1}_{\text{opt}}} 
\newcommand{\CC}[1]			{{#1}^{*}} 
\newcommand{\Op}[1]     {\Mat{#1}} 
\newcommand{\mini}[1] {{\mbox{argmin}}_{#1} \: \: } 
\newcommand{\argmin}[1] {\underset{{#1}}{\mathop{\rm argmin}} \: \: } 
\newcommand{\argmax}[1] {\underset{{#1}}{\mathop{\rm argmax}} \: \: } 
\newcommand{\minimize}{\mathop{\rm minimize} \: \:}
\newcommand{\minimizeu}[1]{\underset{{#1}}{\mathop{\rm minimize}} \: }
\newcommand{\grad}      {\nabla}
\newcommand{\kron}{\otimes} 

\newcommand{\gradt}     {\grad_\z}
\newcommand{\gradx}     {\grad_\x}
\newcommand{\Drv}     	{\Mat{D}} 
\newcommand{\step}      {\text{\textbf{step}}}
\newcommand{\prox}[1]   {\mathbf{prox}_{#1}}
\newcommand{\ind}[1]    {\operatorname{ind}_{#1}}
\newcommand{\proj}[1]   {\Pi_{#1}}
\newcommand{\pointmult}{\odot} 
\newcommand{\rr}   {\mathcal{R}}

\newcommand{\Basis}{\Mat{D}}         		
\newcommand{\Corr}{\Mat{C}}             
\newcommand{\conv}{\ast} 
\newcommand{\meas}{\Vect{b}}            
\newcommand{\Img}{I}                    
\newcommand{\img}{\Vect{i}}             
\newcommand{\vv}{\Vect{v}}
\newcommand{\p}{\Vect{p}}
\newcommand{\Splitvar}{T}                
\newcommand{\splitvar}{\Vect{t}}         
\newcommand{\Splitbasis}{J}                
\newcommand{\splitbasis}{\Vect{j}}         
\newcommand{\var}{\Vari{z}}

\newcommand{\FT}[1]			{\mathcal{F}\left( {#1} \right)} 
\newcommand{\IFT}[1]			{\mathcal{F}^{-1}\left( {#1} \right)} 

\newcommand{\func}{f}
\newcommand{\fMat}{\Mat{K}}

\newcommand{\avar}{\Vari{v}}
\newcommand{\aspvar}{\Vari{z}}

\newcommand{\mask}{\Mat{M}}

\newcommand{\Pen}      		{F} 
\newcommand{\cardset}     {\mathcal{C}}
\newcommand{\Dat}      		{G} 
\newcommand{\Reg}      		{\Gamma} 

\newcommand{\Trans}{\mathbf{\uppercase{T}}} 
\newcommand{\Ph}{\mathbf{\uppercase{\Phi}}} 

\newcommand{\Tvec}{\Vect{T}} 
\newcommand{\Bvec}{\Vect{B}} 

\newcommand{\Wt}{\Mat{W}} 

\newcommand{\Perm}{\Mat{P}} 
\newcommand{\Cblur}{\Mat{C}} 

\newcommand{\DiagFactor}[1]     {\Mat{O}_{ #1 }}  

\newcommand{\Proj}{\Mat{P}}             

\newcommand{\Vector}[1]{\mathbf{#1}}
\newcommand{\Matrix}[1]{\mathbf{#1}}
\newcommand{\Tensor}[1]{\boldsymbol{\mathscr{#1}}}
\newcommand{\TensorUF}[2]{\Matrix{#1}_{(#2)}}

\newcommand{\MatrixKP}[1]{\Matrix{#1}_{\otimes}}
\newcommand{\MatrixKPN}[2]{\Matrix{#1}_{\otimes}^{#2}}

\newcommand{\MatrixKRP}[1]{\Matrix{#1}_{\odot}}
\newcommand{\MatrixKRPN}[2]{\Matrix{#1}_{\odot}^{#2}}

\newcommand{\HP}{\circ}
\newcommand{\HD}{\oslash}

\newcommand{\leftDB}{\left[ \! \left[}
\newcommand{\rightDB}{\right] \! \right]}

\newcommand{\transpose}{T}

\newcommand*\sstrut[1]{\vrule width0pt height0pt depth#1\relax}

\newcommand{\inlineeqnum}{\refstepcounter{equation}~~\mbox{(\theequation)}}
\newcommand{\eqname}[1]{\tag*{#1~(\theequation)}\refstepcounter{equation}}

\newcommand{\lambdas}{\boldsymbol{\lambda}}
\newcommand{\alb}{\boldsymbol{\alpha}} 	
\newcommand{\depth}{\boldsymbol{z}} 	
\newcommand{\albi}{\alpha} 	
\newcommand{\depthi}{z} 	
\newcommand{\ambient}{s}
\newcommand{\jitter}{w}
\newcommand{\z}{\Vect{z}} 							
\newcommand{\x}{\Vect{x}}             	
\newcommand{\y}{\Vect{y}}             	
\newcommand{\Kvar}{\Mat{K}}
\newcommand{\lagrangemult}{\boldsymbol{\nu}}
\newcommand{\scaledlagrange}{\Vect{u}}
\newcommand{\eps}{\epsilon}
\newcommand{\vp}{\Vect{v}}

\begin{affiliations}
 \item Department of Computer Science, Princeton University, Princeton, NJ, USA
 \item Reality Labs Research, Meta, Redmond, WA, USA
 \item Department of Computer Science and Engineering, Pohang University of Science and Technology (POSTECH), Pohang, Republic of Korea
 \item Visual Computing Center, King Abdullah University of Science and Technology (KAUST), Thuwal, Saudi Arabia
 \item [$\dagger$] Corresponding author. E-mail: fheide@princeton.edu
\end{affiliations}
\begin{abstract}
Holographic displays can generate light fields by dynamically modulating the wavefront of a coherent beam of light using a spatial light modulator, promising rich virtual and augmented reality applications. However, the limited spatial resolution of existing dynamic spatial light modulators imposes a tight bound on the diffraction angle. As a result, modern holographic displays possess low \'{e}tendue, which is the product of the display area and the maximum solid angle of diffracted light. The low \'{e}tendue forces a sacrifice of either the field-of-view (FOV) or the display size. In this work, we lift this limitation by presenting neural \'{e}tendue expanders. This new breed of optical elements, which is learned from a natural image dataset, enables higher diffraction angles for ultra-wide FOV while maintaining both a compact form factor and the fidelity of displayed contents to human viewers. With neural \'{e}tendue expanders, we experimentally achieve 64$\times$ \'{e}tendue expansion of natural images in full color, expanding the FOV by an order of magnitude horizontally and vertically, with high-fidelity reconstruction quality (measured in PSNR) over 29\,dB on retinal-resolution images.
\end{abstract}



Holography is the science of creating vivid scenery through carefully crafted interference patterns. This discipline has applications across domains, especially in virtual and augmented reality devices\cite{shi2021towards,itoh2021towards}.
While static holograms can be generated with a suitable recording medium, modern holographic displays typically employ spatial light modulators (SLM) that dynamically modulate the wavefront of a coherent beam\cite{Dorrah2023LightSF,wakunami2016projection}.
However, despite being the workhorse of holography, SLMs suffer from small diffraction angles caused by limitations of modern liquid crystal on silicon (LCoS) technology. Achieving dynamic control with LCoS induces several engineering challenges (e.g., display bandwidth, pixel cross-talk, power consumption), which have imposed a practical lower bound on the pixel pitch\cite{Moser2019Crosstalk}.
Consequently, the \'{e}tendue of holographic displays, which is the product of the field-of-view (FOV) and the display size, is fundamentally limited when maintaining a compact SLM size\cite{lohmann1996space}.
Holographic displays have to trade off FOV for display size, or vice versa, although both are critical for most display applications. For immersive virtual/augmented reality (VR/AR) devices, an FOV of at least $120^\circ$ and an eyebox size greater than $10 \times 10\,\text{mm}^2$ is desired\cite{kuo2020expansion}, where the eyebox or display size is defined as the region within which the eye must reside within to view a hologram. To reach the \'{e}tendue needed for these specifications requires over one billion SLM pixels, which is two orders of magnitude more than what today's LCoS technology achieves\cite{kuo2020expansion}. Manufacturing such a display and dynamically controlling it is beyond modern fabrication and computational capabilities.

Several methods have been proposed to circumvent this problem, including dynamic feedback in the form of eye tracking\cite{choi2020holographic}, spatial integration with multiple SLMs\cite{hahn2008wide}, and temporal integration with laser arrays\cite{lee2020wide} or fast switching digital micromirror devices\cite{Kim2018ExpandedEH,Hellman2019AngularAS}.
However, these approaches require additional dynamic components resulting in high complexity, large form factors, precise timing constraints, and incur additional power consumption. Timing is especially critical for eye tracking solutions as poor latency can result in motion sickness\cite{Wetzstein2010OpticalIP}.
Rewritable photopolymers are promising alternatives to SLMs for holographic displays, however, they are limited by low refresh rates\cite{blanche2010holographic,tay2008updatable}.
Microelectromechanical systems have low pixel counts and bit depth\cite{ersumo2020micromirror}. Researchers have also proposed to make trade-offs within the limited \'{e}tendue of existing holographic displays, such as trading spatial resolution for depth resolution\cite{Dorrah2023LightSF} and optimizing for coherent versus incoherent interference\cite{Yang2022DiffractionengineeredHB}, but these methods do not change the total \'{e}tendue of the display.

Instead, researchers have explored expanding the display \'{e}tendue by employing optical elements with randomized scattering properties in front of an SLM\cite{buckley2006viewingAngle,Huang2015UltrahighcapacityNP,Yu2017UltrahighdefinitionD3,Park2019UltrathinWL,kuo2020expansion,Yu2023Ultrahigh}.
The static nature of these elements facilitates fabrication of pixel areas at the micron-scale, an order of magnitude lower than for a dynamic element such as the SLM, thus resulting in an enlarged diffraction angle\cite{Genet2007LightIT}.
However, existing elements of this type exhibit randomized scattering that is agnostic to the optical setup and the images to be displayed. As modern SLMs have limited degrees of freedom for wavefront shaping, the random modulation delivered by these scattering elements results in low-fidelity \'{e}tendue expanded holograms.
Relaxing high frequencies beyond the human retinal resolution improves the fidelity, however the displayed holograms still suffer from low reconstruction fidelity\cite{kuo2020expansion}.
Furthermore, extensive calibration is necessary in the case where the scattering properties are unknown\cite{mosk2012controlling,conkey2015super,vellekoop2010exploiting,popoff2010measuring,Yu2023Ultrahigh}. Recent work explores using lenses and lenslet arrays\cite{monin2022analyzing} to expand the field-of-view but this approach forces the eye pupil's position to match that of the lenses or lenslets, thus shrinking the effective eyebox size, see Supplementary Note 5 for comparison experiments. Another line of work investigates tilting cascades\cite{monin2022exponential}, however, this system has a large physical footprint consisting of several 4F relays. Specifically, this system requires one 4F relay per 2$\times$ factor of \'{e}tendue expansion, thus three 4F relays are needed to achieve 8$\times$ expansion along one axis, spanning roughly a meter in length. Moreover, the approach requires two parallel cascades to expand the \'{e}tendue in both the horizontal and vertical directions, further increasing the physical form factor.

In this work, we lift these limitations with neural \'{e}tendue expanders, a new breed of static optical elements that have been optimized for \'{e}tendue expansion and accurate reproduction of natural images when combined with an SLM (Fig.~\ref{fig:training}). These optical elements inherit the aforementioned benefits of scattering elements. However, unlike existing random scattering masks, neural \'{e}tendue expanders are jointly learned together with the SLM pattern across a natural image dataset. We devise a differentiable holographic image formation model that enables learning via first-order stochastic optimization. The resulting learned wavefront modulation pushes reconstruction noise outside of the perceivable frequency bands of human visual systems while retaining perceptually critical frequency bands of natural images. We provide analysis that validates that the learned optical elements possess these properties. In simulation, we demonstrate \'{e}tendue-expanded holograms at 64$\times$ \'{e}tendue expansion factor with perceptual display quality over 29\,dB peak signal-to-noise ratio (PSNR), more than one order of magnitude in reduced error over the existing methods.
Furthermore, the expanders also facilitate high-fidelity \'{e}tendue expanded 3D color holograms of natural scenes, see Supplementary Information, and the the elements are also robust to varying pupil positions within the eyebox.
We experimentally validate the method with a 1K-pixel SLM to expand the FOV by $8\times$ in each direction.
Since the improvement in display quality depends on the \'{e}tendue expansion factor and not on the native resolution of the SLM, our 64$\times$ neural \'{e}tendue expander enables high-fidelity ultra-wide-angle holographic projection of natural images when using an 8K-pixel SLM\cite{Sterling2008JVC}. Notably, an 8K-pixel SLM combined with 64$\times$ \'{e}tendue expansion can cover 85\% of the human stereo FOV\cite{Wheelwright2018FOV} with an 18.5\,mm eyebox size, providing a high-fidelity, immersive VR/AR experience.
\section*{Neural \'{e}tendue expansion}
\begin{figure*}[t]
	\centering
	\includegraphics[scale=0.4]{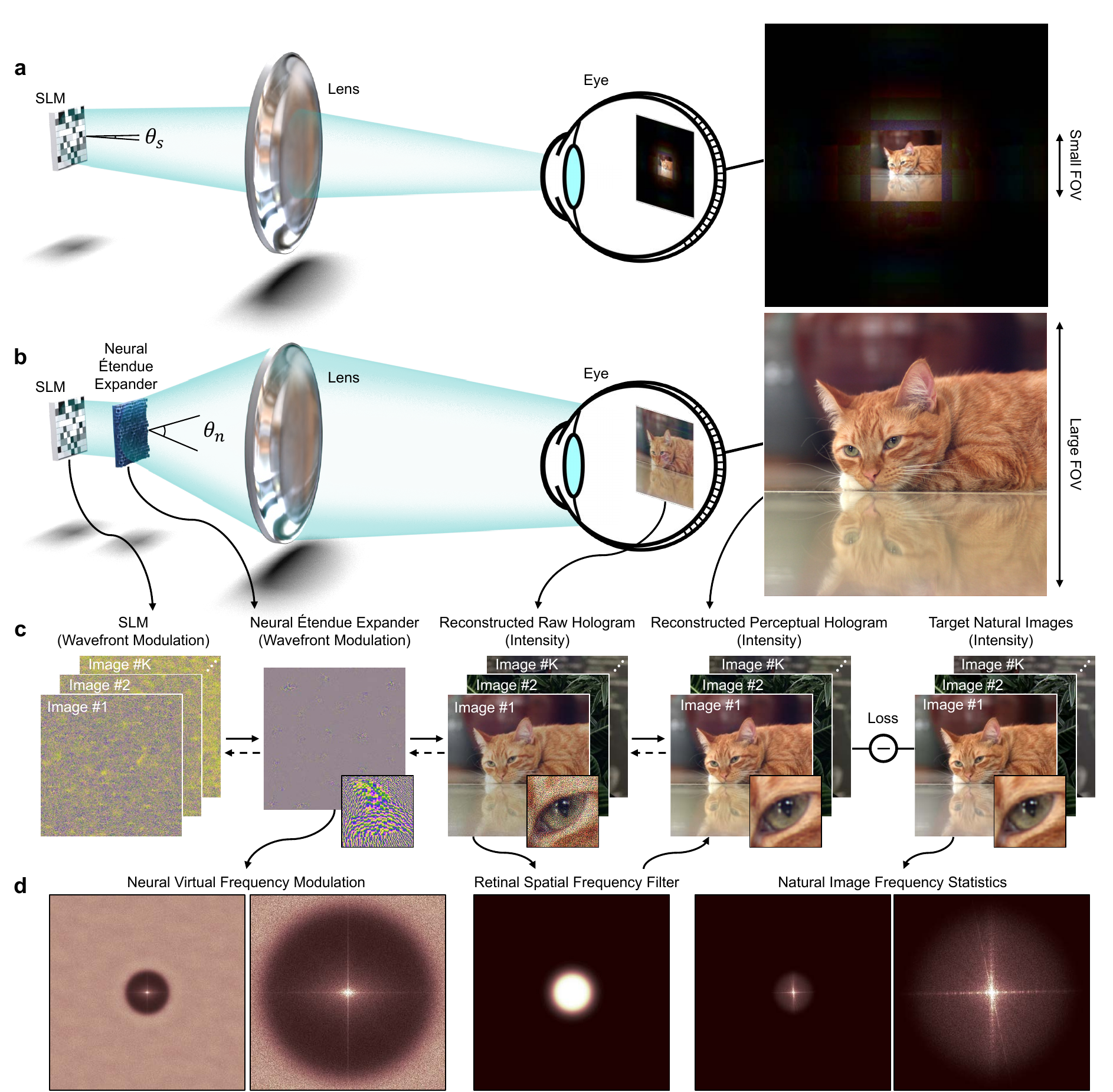}
    \caption{
    \textbf{Neural \'{e}tendue expansion for ultra-wide angle, high-fidelity holograms.}
    \textbf{a} Conventional holographic displays suffer from low \'{e}tendue, resulting in either small FOV or eyebox size. Here, we illustrate a small FOV as $\theta_s$.
    \textbf{b} Introducing a neural \'{e}tendue expander into the display facilitates ultra-wide angle holograms, here we illustrate the increase in FOV as $\theta_n$.
    \textbf{c} We design the neural \'{e}tendue expanders via an end-to-end optimization algorithm that considers the SLM wavefront modulation and the human viewer's perceptual response. One SLM pattern is optimized for each training sample, while the neural \'{e}tendue expander learns a general structure that facilitates hologram generation of any natural image. \textbf{d} The learned neural \'{e}tendue expander preserves the major frequency bands of natural images within the frequency cutoff determined by the resolution of the human retina.
    }
	\label{fig:training}
\end{figure*}

Holographic displays modulate the wavefront of a coherent light beam using an SLM to form an image at a target location.
The \'{e}tendue of a holographic display\cite{chaves2008introduction} is then defined as the product of the SLM display area $A$ and the solid angle of the diffracted light as
\begin{equation}\label{eq:etendue}
G_s=4A\sin^2 \theta_s,
\end{equation}
where $\theta_s=\sin^{-1} \frac{\lambda}{2\Delta_s}$ is the maximum diffraction angle of the SLM, $\lambda$ is the wavelength of light, and $\Delta_s$ is the SLM pixel pitch. Most SLMs have large $\Delta_s$ resulting in small $\theta_s$ as shown in Fig.~\ref{fig:training}a. We enlarge the display \'{e}tendue by placing a neural \'{e}tendue expander in front of the SLM as a static optical element with pixel pitch $\Delta_n < \Delta_s$, see Fig.~\ref{fig:training}b.
The smaller pixel pitch $\Delta_n$ increases the maximum diffraction angle $\theta_n$, resulting in an expanded \'{e}tendue $G_n=4A\sin^2 \theta_n$.

To generate \'{e}tendue-expanded high-fidelity holograms, we propose a computational inverse-design method that learns the wavefront modulation of the neural \'{e}tendue expander by treating it as a layer of trainable neurons that are taught to minimize a loss placed on the formed holographic image, see Fig.~\ref{fig:training}c.
Specifically, we model the holographic image formation in a fully differentiable manner following Fourier optics. We relate the displayed holographic image $I$ to the wavefront modulation of the neural \'{e}tendue expander $\mathcal{E}$ as
\begin{equation}\label{eq:holographic_s}
    I = \left|\mathcal{F}(\mathcal{E}\odot {U(\mathcal{S})})\right|^2,
\end{equation}
where $\mathcal{F}$ is the 2D Fourier transform, $\mathcal{S}$ is the SLM modulation, $U(\cdot)$ is zeroth-order upsampling operator from the low-resolution SLM to the high-resolution neural \'{e}tendue expander, and $\odot$ is the Hadamard product.

The differentiability of Eq.~\eqref{eq:holographic_s} with respect to the modulation variables $\mathcal{E}$ and $\mathcal{S}$ allows us to learn the optimal wavefront modulation of the neural \'{e}tendue expander $\mathcal{E}$ by jointly optimizing the static neural \'{e}tendue expander in concunction with the dynamic SLM modulation patterns $\mathcal{S}$.
That is, for a given image, we optimize the optimal SLM pattern similar to conventional computer-generated holography\cite{gerchberg1972practical,georgiou2008aspects,chakravarthula2019wirtinger}, however, we also simultaneously optimize the neural \'{e}tendue expander.
The SLM and the neural \'{e}tendue expander cooperate to generate an \'{e}tendue-expanded high-fidelity hologram.
We formulate this joint optimization as
\begin{equation}\label{eq:opt}
\mathop \mathrm{minimize} \limits_{\mathcal{E},\mathcal{S}_{\{ 1,...,K\}}} \sum\limits_{k = 1}^K {\left\| {\left( {|{\mathcal{F}}\left( {{\mathcal{ E}} \odot U\left( {{\mathcal{S}_k}} \right)} \right){|^2} - {T_k}} \right)*f} \right\|_2^2} ,
\end{equation}
where $\mathcal{S}_k$ is the SLM wavefront modulation for the $k$-th target image $T_k$ in a natural-image dataset with $K$ training samples,
$*$ is the convolution operator, and $f$ is the low-pass Butterworth filter for approximating the viewer's retinal resolution as a frequency-cutoff function\cite{kuo2020expansion} as
\begin{equation}\label{eq:filter}
f=\mathcal{F}^{-1}\left(\left(1+\left(\frac{\left\|w\right\|^2}{c^2}\right)^5\right)^{-1}\right),
\end{equation}
where $\mathcal{F}^{-1}$ is the inverse 2D Fourier transform, $w$ is the spatial frequency, and $c$ is the cutoff frequency. In order to set the cutoff frequency to be beyond human perceptibility, it suffices to set $c$ to be $\frac{\Delta_n N}{\sqrt{\pi}}$ where $N$ is the SLM pixel count. This is because an 8K-pixel SLM\cite{Sterling2008JVC} can provide an angular resolution of 61 pixels/degree at an eyebox size of 18.5\,mm, see Supplementary Note 3 for details, whereas the angular resolution of the human eye is limited to 60 pixels/degree\cite{kuo2020expansion}.

The optimization objective in Eq.~\eqref{eq:opt} jointly optimizes a single static element $\mathcal{E}$ and a set of SLM patterns $\mathcal{S}_{\{ 1,...,K\}}$ so that the set of generated holograms matches the target set of natural images ${T}_{\{ 1,...,K\}}$.
This objective function is fully differentiable with respect to the wavefront modulations of the SLM and the neural \'{e}tendue expander. As such, we can solve this optimization problem by training the neural \'{e}tendue  expander and the SLM states akin to a shallow neural network by using stochastic gradient solvers\cite{kingma2015method}.
Our computational design approach is data-driven and requires a dataset of natural images.
We used 105 high-resolution training images of natural scenes.
For testing, we used 20 natural images. We use grayscale images when designing neural \'{e}tendue expanders for a monochromatic display, while the original RGB images are used when designing for a trichromatic display.
\section*{Neural \'{e}tendue expanded holographic display}
\begin{figure*}
	\centering
	\includegraphics[scale=0.6]{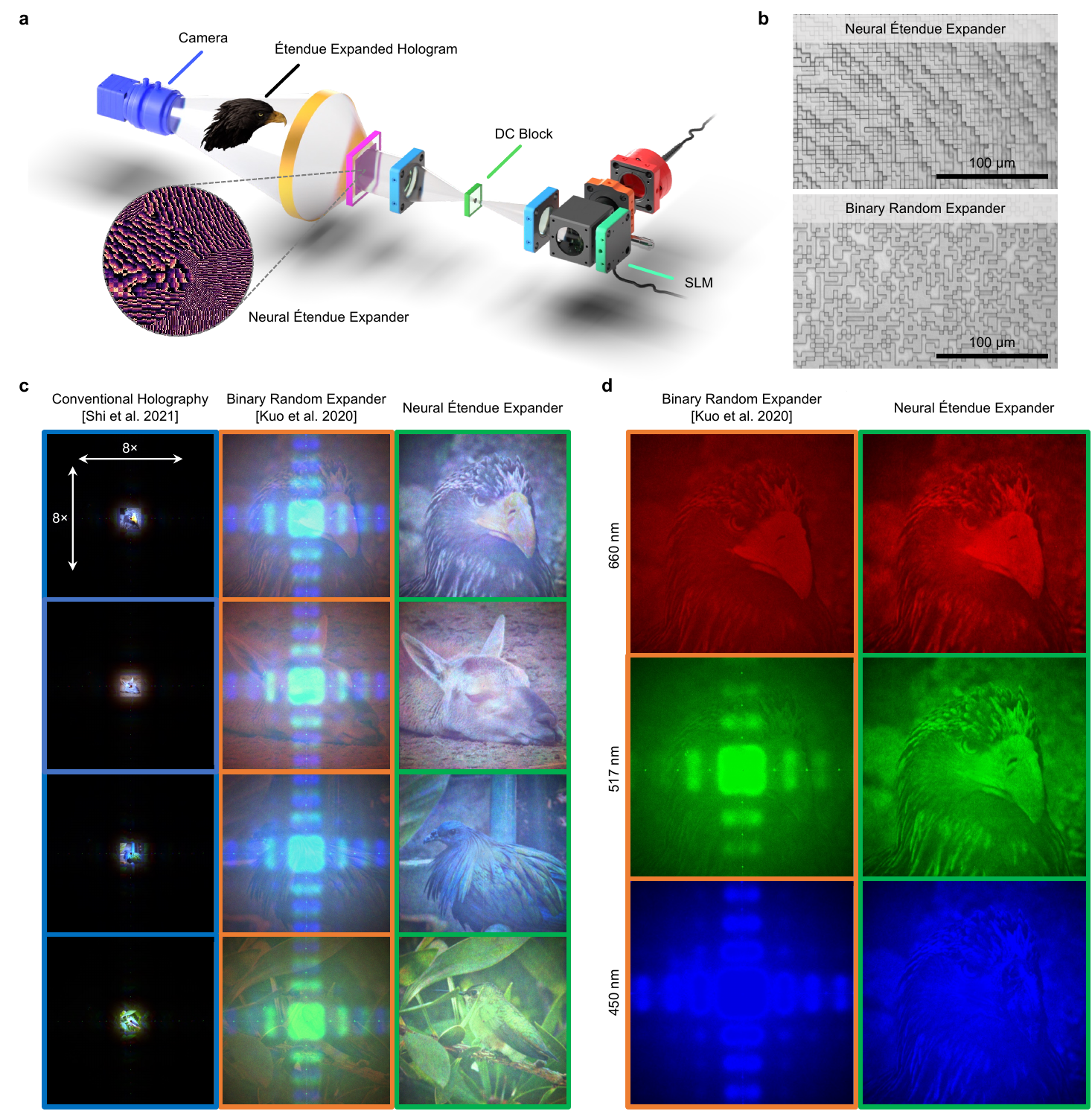}
	\caption{
    \textbf{Experimental demonstration of neural \'{e}tendue expansion.}
    \textbf{a} Schematic of holographic display prototype with the neural \'{e}tendue expander inserted at the conjugate plane of the SLM.
    \textbf{b} Microscope images of a fabricated neural \'{e}tendue expander and a binary random expander\cite{kuo2020expansion}.
    \textbf{c} Captures of holograms generated with the display prototype. The small dark circle in the center of the pictures corresponds to the DC block. Left: Non-\'{e}tendue expanded holograms produced with conventional holography\cite{shi2021towards}. These holograms have extremely low FOV. Middle: $64\times$ \'{e}tendue expanded holograms produced with the binary random expander show low contrast and chromatic artifacts. Right: $64\times$ \'{e}tendue expanded holograms produced with the neural \'{e}tendue expander show high fidelity.
    \textbf{d} Decomposition of $64\times$ \'{e}tendue expanded holograms into constituent colors ($\SI{450}{nm}$, $\SI{517}{nm}$, $\SI{660}{nm}$). We observe improved hologram contrast and less scatter with neural \'{e}tendue expansion, even at the wavelength of $\SI{660}{nm}$ which was used to design the binary random expander.
    }
    \label{fig:prototype}
\end{figure*}

We validate neural \'{e}tendue expansion experimentally with a holographic display prototype. See Fig.~\ref{fig:prototype}a for a schematic of the hardware prototype and Supplementary Notes 9 and 10 for further details on the experimental setup.
We fabricate neural \'{e}tendue expanders with a pitch of $\SI{2}{\micro m}$ with resin stamping, see Supplementary Note 8 for fabrication details.
The fabricated expanders are then placed at the conjugate plane of the SLM to establish pixel-wise correspondence between the SLM and the expander.
A DC block is further employed to filter out the undiffracted light from the SLM.
To assess the proposed elements, we also compare to fabricated binary random expanders\cite{kuo2020expansion} designed for $\SI{660}{nm}$. Microscope images of both expanders are shown in Fig.~\ref{fig:prototype}b.
We acquire holograms corresponding to conventional non-\'{e}tendue expanded holography\cite{shi2021towards}, $64\times$ \'{e}tendue expanded full color holograms produced with the binary random expanders, and $64\times$ \'{e}tendue expanded full color holograms produced with the neural \'{e}tendue expanders. The illumination wavelengths are $\SI{450}{nm}$, $\SI{520}{nm}$, and $\SI{660}{nm}$. We report captures in Fig.~\ref{fig:prototype}c and provide additional measurements in Supplementary Video 1 and in Supplementary Note 1. The captured holograms are tone-mapped for visualization. For fair comparison we applied the same tone-mapping scheme to all holograms, see Supplementary Note 1 for details.

The experimental findings on the display prototype verify that conventional non-\'{e}tendue expanded holography can produce high-fidelity content but at the cost of a small FOV. Increasing the \'{e}tendue via a binary random expander will increase the FOV but at the cost of low image fidelity, even at the design wavelength of $\SI{660}{nm}$, and chromatic artifacts. The \'{e}tendue expanded holograms produced with the neural \'{e}tendue expanders are the only holograms that showcase both ultra-wide-FOV and high-fidelity. The captured holograms demonstrate high contrast and are free from chromatic aberrations. Fig.~\ref{fig:prototype}d reports the \'{e}tendue expanded hologram produced with both expanders at each color wavelength. Since the binary random expander is, by design, only tailored to a single wavelength, in this case $\SI{660}{nm}$, the \'{e}tendue expanded holograms that are generated with it exhibit severe chromatic artifacts. In contrast, holograms generated with neural \'{e}tendue expansion show consistent high-fidelity performance at all illumination wavelengths. Notably, even at the wavelength of $\SI{660}{nm}$ the hologram fidelity is higher for the holograms generated with neural \'{e}tendue expansion, see Fig.~\ref{fig:prototype}c. For comparisons against a uniform random expander, where the phase profile is uniformly randomly selected from within $[0,2\pi]$, see Fig.~\ref{fig:simulation} and Supplementary Note 2.

While our experimental prototype was built for a HOLOEYE-PLUTO which possesses a 1K-pixel resolution, corresponding to a 1\,mm eyebox with $75.6^\circ$ horizontal and vertical FOV, the improvement in hologram fidelity persists across resolutions. Irrespective of the resolution of the SLM, performing $4\times$, $16\times$, or $64\times$ \'{e}tendue expansion with neural \'{e}tendue expanders results in a similar margin of improvement over uniform and binary random expanders. This is because the improvement in fidelity depends only on the \'{e}tendue expansion factor. To validate this we simulate an 8K-pixel SLM with $64\times$ \'{e}tendue expansion and we verify that the improvement in fidelity is maintained. See Supplementary Note 6 for results and further details. Thus, neural \'{e}tendue expansion enables high fidelity expansion for $64\times$ \'{e}tendue expansion for 8K-pixel SLMs\cite{Sterling2008JVC}, providing \'{e}tendue to cover 85\% of the human stereo FOV\cite{Wheelwright2018FOV} with a 18.5\,mm eyebox size, see Supplementary Note 3 for details.
\section*{Characterization of \'{e}tendue expansion}
Next, we analyze the expansion of  \'{e}tendue achieved with the proposed technique. To this end, suppose we want to generate the \'{e}tendue-expanded hologram of only a single scene.
Then, the optimal complex wavefront modulation for the neural \'{e}tendue expander would be the inverse Fourier transform of the target scene, and, as such, we do not require any additional modulation on the SLM. The SLM therefore can be set to zero-phase modulation.
If we generalize this single-image case to diverse natural images, the neural \'{e}tendue expander is expected to preserve the common frequency statistics of natural images, while the SLM fills in the image-specific residual frequencies to generate a specific target image.
In contrast, existing random scatters used for \'{e}tendue expansion do not consider any natural-image statistics\cite{Yu2017UltrahighdefinitionD3,Park2019UltrathinWL,buckley2006viewingAngle,kuo2020expansion}.

To assess whether the optimized neural \'{e}tendue expander $\mathcal{E}$, shown in Fig.~\ref{fig:training}b, has learned the image statistics of the training set we evaluate the virtual frequency modulation $\widetilde{\mathcal{E}}$, defined as the spectrum of the generated image with the neural \'{e}tendue expander and the zero-phase SLM modulation as
\begin{equation}\label{eq:virtual}
\widetilde{\mathcal{E}} = \mathcal{F}\left(|\mathcal{F}(\mathcal{E})|^2\right).
\end{equation}
The findings in Fig.~\ref{fig:training}d confirm that the magnitude of the virtual frequency modulation $|\widetilde{\mathcal{E}}|$ resembles the magnitude spectrum of natural images within the passband.
Moreover, we observe that the virtual frequency modulation pushes undesirable energy outside of the passband of the human retina as imperceptible high-frequency noise.

To further understand this property of a neural \'{e}tendue expander, we consider the reconstruction loss $\mathcal{L}_T$ for a specific target image $T$.
Using the zero-phase setting for the SLM as an initial point for the first-order stochastic optimization and applying Parseval's theorem places an upper bound on the reconstruction loss
\begin{equation}\label{eq:proof2}
\mathcal{L}_T = \min \limits_{\mathcal{S}} \left\| {\left( {|{\mathcal{F}}\left( {{\mathcal{ E}} \odot U\left( {{\mathcal{S}}} \right)} \right){|^2} - {T}} \right)*f} \right\|_2^2 \leq \frac{1}{N}\left\| {\left( {\widetilde{\mathcal{E}} - \mathcal{F}(T)} \right) \odot \mathcal{F}(f)} \right\|_2^2,
\end{equation}
where $N$ is the pixel count of the neural \'{e}tendue expander. Please see Supplementary Note 3 for further details of how this upper bound is found.
Therefore, obtaining the optimal neural \'{e}tendue expander, which minimizes the reconstruction loss $\mathcal{L}_T$, results in the virtual frequency modulation $\widetilde{\mathcal{E}}$ that resembles the natural-image spectrum $\mathcal{F}(T)$ averaged over diverse natural images. Also, the retinal frequency filter $\mathcal{F}(f)$ leaves the higher spectral bands outside of the human retinal resolution unconstrained. This allows the neural \'{e}tendue expander to push undesirable energy towards higher frequency bands, which then manifests as imperceptible high-frequency noise to human viewers.

\begin{figure*}
	\centering
	\includegraphics[scale=0.4]{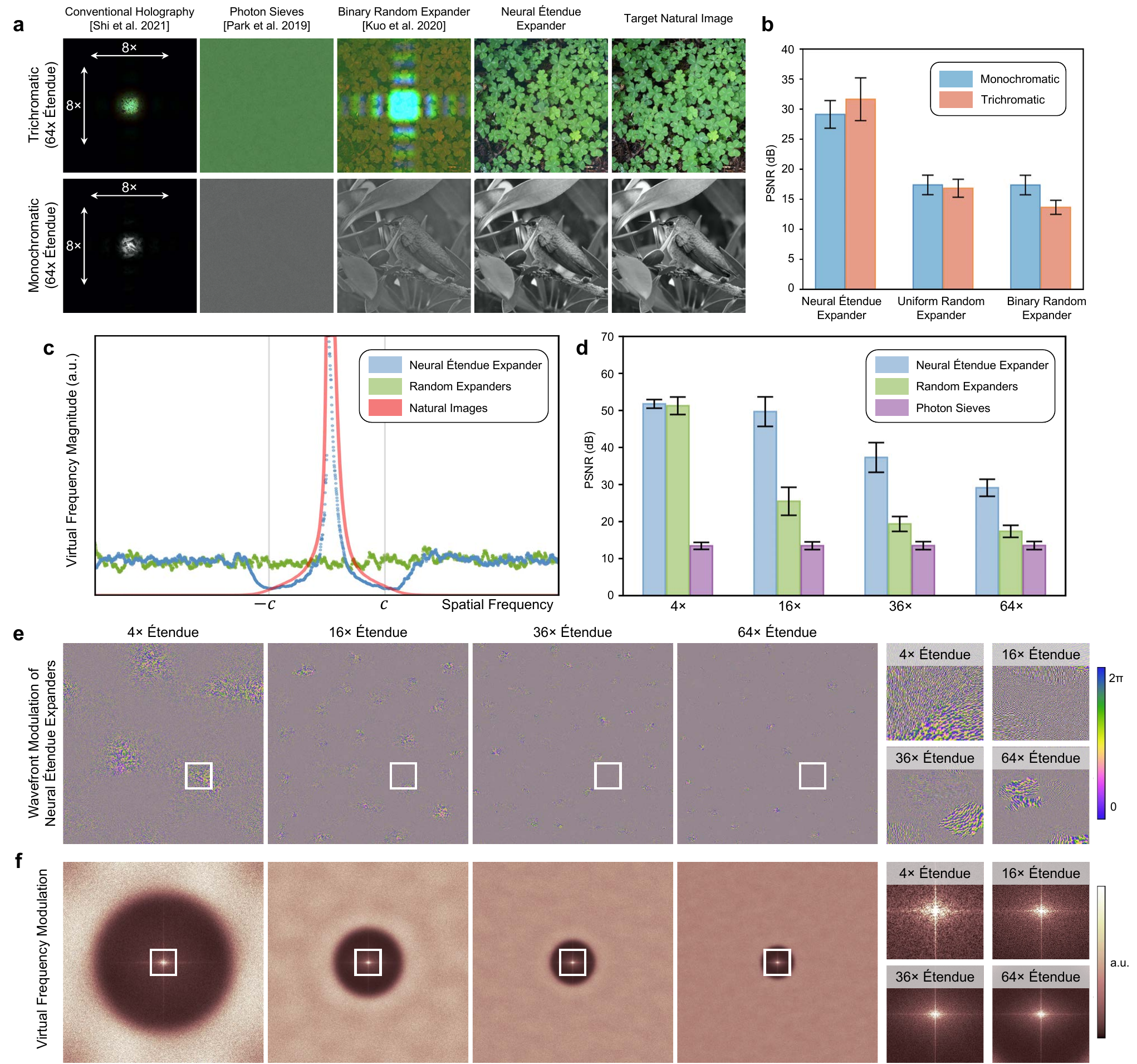}
	\caption{
    \textbf{\'{E}tendue expander characterization.}
    \textbf{a} The 64$\times$ \'{e}tendue expanded holograms generated with neural \'{e}tendue expanders have the highest fidelity with respect to the target natural image, for both the trichromatic and monochromatic cases. In comparison, the holograms generated with binary random expanders\cite{kuo2020expansion} or photon sieves\cite{Park2019UltrathinWL} show lower contrast and more speckle noise. Photon sieves could generate \'{e}tendue expanded holograms of sparse points but not of natural scenes. A low \'{e}tendue hologram generated with conventional holography\cite{shi2021towards} and no expander is included for comparison.
    \textbf{b} Quantitative performance comparison of $64\times$ \'{e}tendue expanded holograms when using neural, uniform random, and binary random expanders. The metrics are evaluated over an unseen test set.
    \textbf{c} Virtual frequency modulation cross section. Neural \'{e}tendue expanders push reconstruction artifacts outside of the perceivable frequency bands of human vision while producing a natural image frequency spectrum within the passband as predicted by Eq.~\ref{eq:proof2}. In contrast, the both uniform and binary random expanders exhibit a flat spectrum which reduces the reconstruction quality within the passband. The cutoff frequency is indicated by $c$.
    \textbf{d} Quantitative reconstruction quality of neural \'{e}tendue expansion, random expansion, and photon sieves\cite{Park2019UltrathinWL} for increasing \'{e}tendue expansion factors for the monochromatic case. Uniform and binary random expansion both achieve the same performance for a single wavelength.
    \textbf{e} Visualization of the learned expanders for increasing \'{e}tendue expansion factors. We observe that the learned modulation structures contain both high and low frequency components.
    \textbf{f} Visualization of the corresponding virtual frequency modulation for each expander.
    \label{fig:simulation}
    }
\end{figure*}

We investigate the image statistics preserved by the neural \'{e}tendue expanders by visualizing $|\widetilde{\mathcal{E}}|$. Fig.~\ref{fig:simulation}e visualizes the learned expander pattern $\mathcal{E}$ for increasing \'{e}tendue expansion factors, specifically for $4\times$, $16\times$, $36\times$, and $64\times$ expansion. Unlike uniform and binary random expanders, the learned expanders exhibit high and low frequency structures. Fig.~\ref{fig:simulation}f shows the corresponding virtual frequency $|\widetilde{\mathcal{E}}|$ for each \'{e}tendue expansion factor. We observe that the interwoven high and low frequency patterns on each learned expander correspond to a virtual frequency that pushes noise outside of the retinal frequency bands defined by $\mathcal{F}(f)$. Furthermore, the frequency structure within the passband resembles the frequency structure of the natural image training dataset, see Fig.~\ref{fig:simulation}c.

To characterize the hologram reconstruction with the proposed neural \'{e}tendue expander we simulate a Fourier holographic setup that has been augmented with a neural \'{e}tendue expander. Fig.~\ref{fig:simulation}a reports qualitative examples of trichromatic and monochromatic reconstructions achieved with neural \'{e}tendue expanders, binary random expanders\cite{kuo2020expansion}, photon sieves\cite{Park2019UltrathinWL}, and conventional holography\cite{shi2021towards}. See Supplementary Note 2 for additional qualitative comparisons and for comparisons against uniform random expanders.
The uniform random expander is constructed by assigning each pixel a phase that is uniformly randomly chosen within $[0,2\pi]$. To ensure at least $2\pi$ phase is available for all wavelengths the $[0,2\pi]$ phase range is defined for $\SI{660}{nm}$. Conventional holography is subject to a low display \'{e}tendue that is limited by the SLM native resolution, thus resulting in a low FOV. Photon sieves, binary random expanders, and uniform random expanders have low reconstruction fidelity, resulting in severe noise and low contrast in the generated holograms. In the case of the trichromatic holograms, both uniform and binary random expanders do not facilitate consistent \'{e}tendue expansion at all wavelengths, which results in chromatic artifacts. Although the uniform random expander provides at least $2\pi$ phase coverage for all wavelengths, the variation in refractive index across wavelengths results in differing phase profiles. Thus, although the uniform random expander has the same degree of quantization as neural \'{e}tendue expanders, it does not enable \'{e}tendue expanded trichromatic holograms. Photon sieves scatter light equally across wavelengths but their randomized amplitude-only modulation does not allow for high-fidelity reconstruction of natural images, see Fig.~\ref{fig:simulation}d for quantitative metrics and Supplementary Note 2 for qualitative examples and additional metrics. Neural \'{e}tendue expansion is the only technique that facilitates high-fidelity reconstructions for both trichromatic and monochromatic setups. We quantitatively verify this by evaluating the reconstruction fidelity on an unseen test dataset, where fidelity is measured in peak signal-to-noise ratio (PSNR). Fig.~\ref{fig:simulation}b shows that neural \'{e}tendue expanders achieve over 14\,dB PSNR improvement favorable to other expanders when generating $64\times$ \'{e}tendue expanded trichromatic holograms. For monochromatic holograms, neural \'{e}tendue expansion achieves over 10\,dB PSNR improvement. Thus, neural \'{e}tendue expansion allows for an order of magnitude improvement over existing \'{e}tendue expansion methods. See Fig.~\ref{fig:simulation}d for quantitative evaluations at different \'{e}tendue expansion factors. See Supplementary Note 2 for further simulation details and more comparison examples.

In addition to field-of-view, we also investigate the eyebox that is produced with neural \'{e}tendue expansion. By initializing the learning process with a uniform random expander we bias the optimized solution towards expanders that distribute energy throughout the eyebox, in contrast to a quadratic phase profiles\cite{monin2022analyzing} that concentrate the energy at fixed points. Thus, the viewer's eye pupil can freely move within the eyebox and observe the wide field-of-view hologram at any location. We incorporate pupil-aware optimization\cite{Chakravarthula2022pupil} to preserve the perceived hologram quality at different eye pupil locations. See Supplementary Note 5 for findings.
Finally, we also investigate 3D \'{e}tendue expanded holograms. We find that neural \'{e}tendue expansion also enables higher fidelity \'{e}tendue expanded 3D color holograms.  We note that existing methods on \'{e}tendue expanded holography has focused on monochromatic 3D holograms\cite{kuo2020expansion,monin2022analyzing,monin2022exponential}. Photon sieves\cite{Park2019UltrathinWL} only achieves 3D color holography for sparse points. See Supplementary Note 4 for a discussion of these findings.
\section*{Discussion}
In this work, we introduce neural \'{e}tendue expanders as an optical element that expands the \'{e}tendue of existing holographic displays without sacrificing displayed hologram fidelity. Neural \'{e}tendue expanders are learned  from a natural image dataset and are jointly optimized with the SLM's wavefront modulation. Akin to a shallow neural network, this new breed of optical elements allows us to tailor the wavefront modulation element to the display of natural images and maximize display quality perceivable by the human eye.
As the first learned optics for \'{e}tendue expansion, we achieve \'{e}tendue expansion factor 64$\times$ with over 29\,dB PSNR reconstructions, an order of magnitude improvement over existing approaches.
This means that expansion factor $64\times$ combined with an 8K-pixel SLM can enable high-fidelity, ultra-wide-angle holographic projection of natural images with $126^\circ$ FOV and 18.5\,mm eyebox size, covering more than 85\% of the human FOV. Furthermore, neural \'{e}tendue expanders support multi-wavelength illumination for color holograms.
The expanders also support 3D color holography and viewer pupil-awareness.
We envision that future holographic displays may incorporate the described optical design approach into their construction, especially for VR/AR displays. Extending our work to utilize other types of emerging optics such as metasurfaces may prove to be a promising direction for future work, as diffraction angles can be greatly enlarged by nano-scale metasurface features\cite{scheuer2015metasurfaces} and additional properties of light such as polarization can be modulated using meta-optics\cite{dorrah2021metasurface}.


\begin{methods}
\subsection{Simulation} We used PyTorch to design and evaluate the neural \'{e}tendue expanders. See Supplementary Notes 2 and 3 for details on the optimization framework, evaluation, and analysis.

\subsection{Fabrication} The expanders are physically realized as diffractive optical elements (DOE). Fabricating the DOEs consists of several stages. The first stage consists of etching the negative of the desired pattern onto a substrate. This etching is performed with laser beam lithography. The etched substrate forms a stamp which is then pressed onto a resin mold that is mounted on a glass substrate. The resin itself contains the final pattern. The resin has a wavelength dependent refractive index that we incorporate into our design framework. For the resin we used, the refractive indices are 1.5081 for $\SI{660}{nm}$, 1.5159 for $\SI{517}{nm}$, and 1.5223 for $\SI{450}{nm}$. See Supplementary Note 8 for details.

\subsection{Experimental Setup}
We evaluated the neural \'{e}tendue expanders using a prototype holographic display. The prototype consists of a HOLOEYE-PLUTO SLM, a 4F system, a DC block, and a camera for imaging the \'{e}tendue expanded holograms. See Supplementary Notes 9 and 10 for details.

\subsection{Data Availability}
The code and data used to generate the findings of this study will be made public on GitHub.

\subsection{Code Availability}
The code and data used to generate the findings of this study will be made public on GitHub.
\end{methods}



\bibliographystyle{naturemag}
\bibliography{reference}


\begin{addendum}
	\item The authors acknowledge the use of Princeton's Imaging and Analysis Center (IAC), which is partially supported by the Princeton Center for Complex Materials (PCCM), a National Science Foundation (NSF) Materials Research Science and Engineering Center (MRSEC; DMR-2011750). We thank Ilya Chugunov for providing natural image data.
	\item [Author Contributions] 
	\item [Competing Interests] The authors declare no competing financial interests.
	\item [Supplementary Information] Supplementary Information accompanies this manuscript as part of the submission files.
	\item [Correspondence] Correspondence should be addressed to F.H.
\end{addendum}

\end{document}


\maketitle

\begin{affiliations}
 \item Department of Computer Science, Princeton University, Princeton, NJ, USA
 \item Reality Labs Research, Meta, Redmond, WA, USA
 \item Department of Computer Science and Engineering, Pohang University of Science and Technology (POSTECH), Pohang, Republic of Korea
 \item Visual Computing Center, King Abdullah University of Science and Technology (KAUST), Thuwal, Saudi Arabia
 \item [$\dagger$] Corresponding author. E-mail: fheide@princeton.edu
\end{affiliations}


\definecolor{brightray}{rgb}{0.8,0.8,0.8}
\definecolor{Gray}{rgb}{0.5,0.5,0.5}
\definecolor{darkblue}{rgb}{0,0,0.7}
\definecolor{orange}{rgb}{1,.5,0} 
\definecolor{red}{rgb}{1,0,0} 
\definecolor{blue}{rgb}{0,0,1} 
\definecolor{darkgreen}{rgb}{0,0.7,0} 
\definecolor{darkred}{rgb}{0.7,0,0} 

\newcommand{\heading}[1]{\noindent\textbf{#1}}
\newcommand{\note}[1]{{{\textcolor{orange}{#1}}}}
\newcommand{\todo}[1]{{\textcolor{red}{TODO: #1}}}
\newcommand{\changed}[1]{{\textcolor{blue}{#1}}}
\newcommand{\removed}[1]{{\textcolor{brightray}{{#1}}}}
\newcommand{\revision}[1]{{{#1}}}
\newcommand{\place}[1]{ \begin{itemize}\item\textcolor{darkblue}{#1}\end{itemize}}
\newcommand{\de}{\mathrm{d}}

\newcommand{\baek}[1]{{\textcolor{darkgreen}{[BAEK: #1]}}}
\newcommand{\ethan}[1]{{\textcolor{darkred}{[ETHAN: #1]}}}
\newcommand{\felix}[1]{{\textcolor{darkblue}{[FELIX: #1]}}}

\newcommand{\BEAS}{\begin{eqnarray*}}
\newcommand{\EEAS}{\end{eqnarray*}}
\newcommand{\BEA}{\begin{eqnarray}}
\newcommand{\EEA}{\end{eqnarray}}
\newcommand{\BEQ}{\begin{equation}}
\newcommand{\EEQ}{\end{equation}}
\newcommand{\BIT}{\begin{itemize}}
\newcommand{\EIT}{\end{itemize}}
\newcommand{\BNUM}{\begin{enumerate}}
\newcommand{\ENUM}{\end{enumerate}}

\newcommand{\BA}{\begin{array}}
\newcommand{\EA}{\end{array}}

\newcommand{\eg}{{\it e.g.}}
\newcommand{\ie}{{\it i.e.}}
\newcommand{\etc}{{\it etc.}}

\newcommand{\ones}{\mathbf 1}

\newcommand{\reals}{{\mbox{\bf R}}}
\newcommand{\integers}{{\mbox{\bf Z}}}
\newcommand{\eqbydef}{\mathrel{\stackrel{\Delta}{=}}}
\newcommand{\complex}{{\mbox{\bf C}}}
\newcommand{\symm}{{\mbox{\bf S}}}  

\newcommand{\Span}{\mbox{\textrm{span}}}
\newcommand{\Range}{\mbox{\textrm{range}}}
\newcommand{\nullspace}{{\mathcal N}}
\newcommand{\range}{{\mathcal R}}
\newcommand{\Nullspace}{\mbox{\textrm{nullspace}}}
\newcommand{\Rank}{\mathop{\bf Rank}}
\newcommand{\Tr}{\mathop{\bf Tr}}
\newcommand{\diag}{\mathop{\bf diag}}
\newcommand{\lambdamax}{{\lambda_{\rm max}}}
\newcommand{\lambdamin}{\lambda_{\rm min}}

\newcommand{\Expect}{\mathop{\bf E{}}}
\newcommand{\Prob}{\mathop{\bf Prob}}
\newcommand{\erf}{\mathop{\bf erf}}

\newcommand{\Co}{{\mathop {\bf Co}}}
\newcommand{\co}{{\mathop {\bf Co}}}
\newcommand{\dist}{\mathop{\bf dist{}}}
\newcommand{\Ltwo}{{\bf L}_2}
\newcommand{\QED}{~~\rule[-1pt]{8pt}{8pt}}\def\qed{\QED}
\newcommand{\approxleq}{\mathrel{\smash{\makebox[0pt][l]{\raisebox{-3.4pt}{\small$\sim$}}}{\raisebox{1.1pt}{$<$}}}}
\newcommand{\epi}{\mathop{\bf epi}}

\newcommand{\vol}{\mathop{\bf vol}}
\newcommand{\Vol}{\mathop{\bf vol}}
\newcommand{\Card}{\mathop{\bf card}}

\newcommand{\dom}{\mathop{\bf dom}}
\newcommand{\aff}{\mathop{\bf aff}}
\newcommand{\cl}{\mathop{\bf cl}}
\newcommand{\Angle}{\mathop{\bf angle}}
\newcommand{\intr}{\mathop{\bf int}}
\newcommand{\relint}{\mathop{\bf rel int}}
\newcommand{\bd}{\mathop{\bf bd}}
\newcommand{\vect}{\mathop{\bf vec}}
\newcommand{\dsp}{\displaystyle}
\newcommand{\foequal}{\simeq}
\newcommand{\VOL}{{\mbox{\bf vol}}}
\newcommand{\xopt}{x^{\rm opt}}

\newcommand{\Xb}{{\mbox{\bf X}}}
\newcommand{\xst}{x^\star}
\newcommand{\varphist}{\varphi^\star}
\newcommand{\lambdast}{\lambda^\star}
\newcommand{\Zst}{Z^\star}
\newcommand{\fstar}{f^\star}
\newcommand{\xstar}{x^\star}
\newcommand{\xc}{x^\star}
\newcommand{\lambdac}{\lambda^\star}
\newcommand{\lambdaopt}{\lambda^{\rm opt}}

\newcommand{\geqK}{\mathrel{\succeq_K}}
\newcommand{\gK}{\mathrel{\succ_K}}
\newcommand{\leqK}{\mathrel{\preceq_K}}
\newcommand{\lK}{\mathrel{\prec_K}}
\newcommand{\geqKst}{\mathrel{\succeq_{K^*}}}
\newcommand{\gKst}{\mathrel{\succ_{K^*}}}
\newcommand{\leqKst}{\mathrel{\preceq_{K^*}}}
\newcommand{\lKst}{\mathrel{\prec_{K^*}}}
\newcommand{\geqL}{\mathrel{\succeq_L}}
\newcommand{\gL}{\mathrel{\succ_L}}
\newcommand{\leqL}{\mathrel{\preceq_L}}
\newcommand{\lL}{\mathrel{\prec_L}}
\newcommand{\geqLst}{\mathrel{\succeq_{L^*}}}
\newcommand{\gLst}{\mathrel{\succ_{L^*}}}
\newcommand{\leqLst}{\mathrel{\preceq_{L^*}}}
\newcommand{\lLst}{\mathrel{\prec_{L^*}}}

\newtheorem{theorem}{Theorem}[section]
\newtheorem{corollary}{Corollary}[theorem]
\newtheorem{lemma}[theorem]{Lemma}
\newtheorem{proposition}[theorem]{Proposition}

\newenvironment{algdesc}%
{\begin{quote}}{\end{quote}}

\def\figbox#1{\framebox[\hsize]{\hfil\parbox{0.9\hsize}{#1}}}

\makeatletter
\long\def\@makecaption#1#2{
   \vskip 9pt
   \begin{small}
   \setbox\@tempboxa\hbox{{\bf #1:} #2}
   \ifdim \wd\@tempboxa > 5.5in
        \begin{center}
        \begin{minipage}[t]{5.5in}
        \addtolength{\baselineskip}{-0.95pt}
        {\bf #1:} #2 \par
        \addtolength{\baselineskip}{0.95pt}
        \end{minipage}
        \end{center}
   \else
    \hbox to\hsize{\hfil\box\@tempboxa\hfil}
   \fi
   \end{small}\par
}
\makeatother

\newcounter{oursection}
\newcommand{\oursection}[1]{
 \addtocounter{oursection}{1}
 \setcounter{equation}{0}
 \clearpage \begin{center} {\Huge\bfseries #1} \end{center}
 {\vspace*{0.15cm} \hrule height.3mm} \bigskip
 \addcontentsline{toc}{section}{#1}
}
\newcommand{\oursectionf}[1]{  
 \addtocounter{oursection}{1}
 \setcounter{equation}{0}
 \foilhead[-.5cm]{#1 \vspace*{0.8cm} \hrule height.3mm }
 \LogoOn
}
\newcommand{\oursectionfl}[1]{  
 \addtocounter{oursection}{1}
 \setcounter{equation}{0}
 \foilhead[-1.0cm]{#1}
 \LogoOn
}

\newcommand{\Mat}[1]    {{\ensuremath{\mathbf{\uppercase{#1}}}}} 
\newcommand{\Vect}[1]   {{\ensuremath{\mathbf{\lowercase{#1}}}}} 
\newcommand{\Vari}[1]   {{\ensuremath{\mathbf{\lowercase{#1}}}}} 
\newcommand{\Id}				{\mathbb{I}} 
\newcommand{\Diag}[1] 	{\operatorname{diag}\left({ #1 }\right)} 
\newcommand{\Opt}[1] 	  {{#1}_{\text{opt}}} 
\newcommand{\CC}[1]			{{#1}^{*}} 
\newcommand{\Op}[1]     {\Mat{#1}} 
\newcommand{\mini}[1] {{\mbox{argmin}}_{#1} \: \: } 
\newcommand{\argmin}[1] {\underset{{#1}}{\mathop{\rm argmin}} \: \: } 
\newcommand{\argmax}[1] {\underset{{#1}}{\mathop{\rm argmax}} \: \: } 
\newcommand{\minimize}{\mathop{\rm minimize} \: \:}
\newcommand{\minimizeu}[1]{\underset{{#1}}{\mathop{\rm minimize}} \: }
\newcommand{\grad}      {\nabla}
\newcommand{\kron}{\otimes} 

\newcommand{\gradt}     {\grad_\z}
\newcommand{\gradx}     {\grad_\x}
\newcommand{\Drv}     	{\Mat{D}} 
\newcommand{\step}      {\text{\textbf{step}}}
\newcommand{\prox}[1]   {\mathbf{prox}_{#1}}
\newcommand{\ind}[1]    {\operatorname{ind}_{#1}}
\newcommand{\proj}[1]   {\Pi_{#1}}
\newcommand{\pointmult}{\odot} 
\newcommand{\rr}   {\mathcal{R}}

\newcommand{\Basis}{\Mat{D}}         		
\newcommand{\Corr}{\Mat{C}}             
\newcommand{\conv}{\ast} 
\newcommand{\meas}{\Vect{b}}            
\newcommand{\Img}{I}                    
\newcommand{\img}{\Vect{i}}             
\newcommand{\vv}{\Vect{v}}
\newcommand{\p}{\Vect{p}}
\newcommand{\Splitvar}{T}                
\newcommand{\splitvar}{\Vect{t}}         
\newcommand{\Splitbasis}{J}                
\newcommand{\splitbasis}{\Vect{j}}         
\newcommand{\var}{\Vari{z}}

\newcommand{\FT}[1]			{\mathcal{F}\left( {#1} \right)} 
\newcommand{\IFT}[1]			{\mathcal{F}^{-1}\left( {#1} \right)} 

\newcommand{\func}{f}
\newcommand{\fMat}{\Mat{K}}

\newcommand{\avar}{\Vari{v}}
\newcommand{\aspvar}{\Vari{z}}

\newcommand{\mask}{\Mat{M}}

\newcommand{\Pen}      		{F} 
\newcommand{\cardset}     {\mathcal{C}}
\newcommand{\Dat}      		{G} 
\newcommand{\Reg}      		{\Gamma} 

\newcommand{\Trans}{\mathbf{\uppercase{T}}} 
\newcommand{\Ph}{\mathbf{\uppercase{\Phi}}} 

\newcommand{\Tvec}{\Vect{T}} 
\newcommand{\Bvec}{\Vect{B}} 

\newcommand{\Wt}{\Mat{W}} 

\newcommand{\Perm}{\Mat{P}} 
\newcommand{\Cblur}{\Mat{C}} 

\newcommand{\DiagFactor}[1]     {\Mat{O}_{ #1 }}  

\newcommand{\Proj}{\Mat{P}}             

\newcommand{\Vector}[1]{\mathbf{#1}}
\newcommand{\Matrix}[1]{\mathbf{#1}}
\newcommand{\Tensor}[1]{\boldsymbol{\mathscr{#1}}}
\newcommand{\TensorUF}[2]{\Matrix{#1}_{(#2)}}

\newcommand{\MatrixKP}[1]{\Matrix{#1}_{\otimes}}
\newcommand{\MatrixKPN}[2]{\Matrix{#1}_{\otimes}^{#2}}

\newcommand{\MatrixKRP}[1]{\Matrix{#1}_{\odot}}
\newcommand{\MatrixKRPN}[2]{\Matrix{#1}_{\odot}^{#2}}

\newcommand{\HP}{\circ}
\newcommand{\HD}{\oslash}

\newcommand{\leftDB}{\left[ \! \left[}
\newcommand{\rightDB}{\right] \! \right]}

\newcommand{\transpose}{T}

\newcommand*\sstrut[1]{\vrule width0pt height0pt depth#1\relax}

\newcommand{\inlineeqnum}{\refstepcounter{equation}~~\mbox{(\theequation)}}
\newcommand{\eqname}[1]{\tag*{#1~(\theequation)}\refstepcounter{equation}}

\newcommand{\lambdas}{\boldsymbol{\lambda}}
\newcommand{\alb}{\boldsymbol{\alpha}} 	
\newcommand{\depth}{\boldsymbol{z}} 	
\newcommand{\albi}{\alpha} 	
\newcommand{\depthi}{z} 	
\newcommand{\ambient}{s}
\newcommand{\jitter}{w}
\newcommand{\z}{\Vect{z}} 							
\newcommand{\x}{\Vect{x}}             	
\newcommand{\y}{\Vect{y}}             	
\newcommand{\Kvar}{\Mat{K}}
\newcommand{\lagrangemult}{\boldsymbol{\nu}}
\newcommand{\scaledlagrange}{\Vect{u}}
\newcommand{\eps}{\epsilon}
\newcommand{\vp}{\Vect{v}}

\noindent In this document we provide additional discussion and results in support of the primary manuscript.

\section*{Supplementary Video 1: Experimental Assessment for Dynamic Scenes}
\label{sec:video}
The Supplementary Videos are experimental video captures of the \'{e}tendue expanded holograms generated with neural \'{e}tendue expanders. For comparison, \'{e}tendue expanded holograms generated with binary random expanders\cite{kuo2020expansion} and non-\'{e}tendue expanded holograms\cite{shi2021towards} are included. All results shown are captured on the same hardware prototype as described in the manuscript. The results shown in the videos are 2D holograms, see Supplementary Note 4 for 3D hologram results. We use time multiplexing CGH where we compute 3 independent holograms for each frame of each video. The 3 holograms are displayed and captured sequentially and each frame of each video is then computed as the average of the 3 holograms. The refresh rate of the HOLOEYE-PLUTO SLM that we used is 60 Hz, resulting in an effective framerate of 20 Hz after time averaging.

\clearpage

\section*{Supplementary Note 1: Additional Experimental Results}
\label{sec:experimental}
\begin{figure*}
	\centering
    	\if\loadFigures0
    \else
    \includegraphics[width=\linewidth]{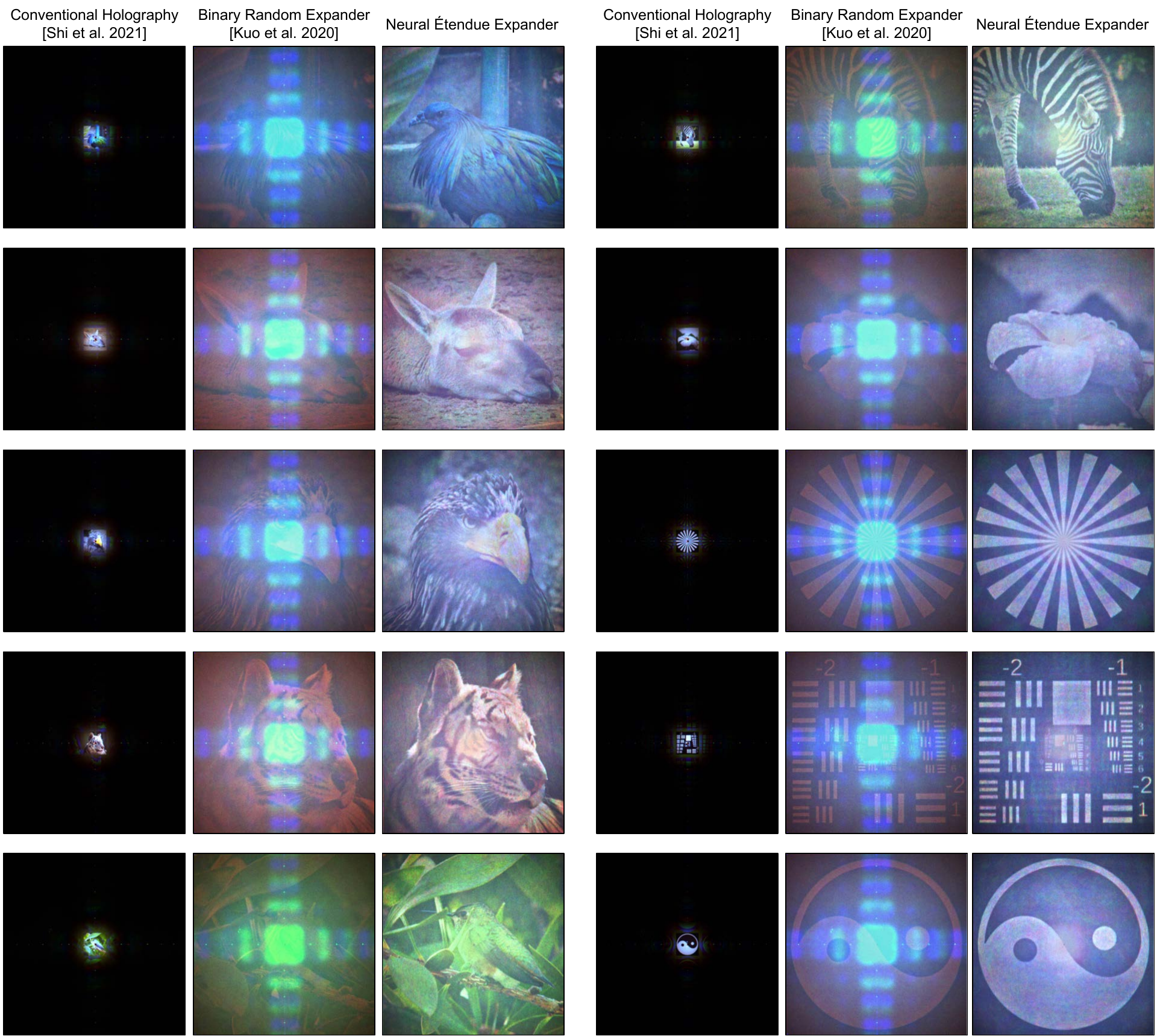}
    \fi
    \caption{\label{fig:real_color_64}
Experimentally captured color holograms at $64\times$ \'{e}tendue expansion. The wavelengths used are $\SI{660}{nm}$, $\SI{517}{nm}$, and $\SI{450}{nm}$. For comparison, \'{e}tendue expanded holograms generated with random expanders\cite{kuo2020expansion} and non-\'{e}tendue expanded holograms\cite{shi2021towards} are included. These results supplement the experimental findings from Fig.~2 of the main manuscript.
}		
\end{figure*}

\begin{figure*}
	\centering
    	\if\loadFigures0
    \else
    \includegraphics[width=\linewidth]{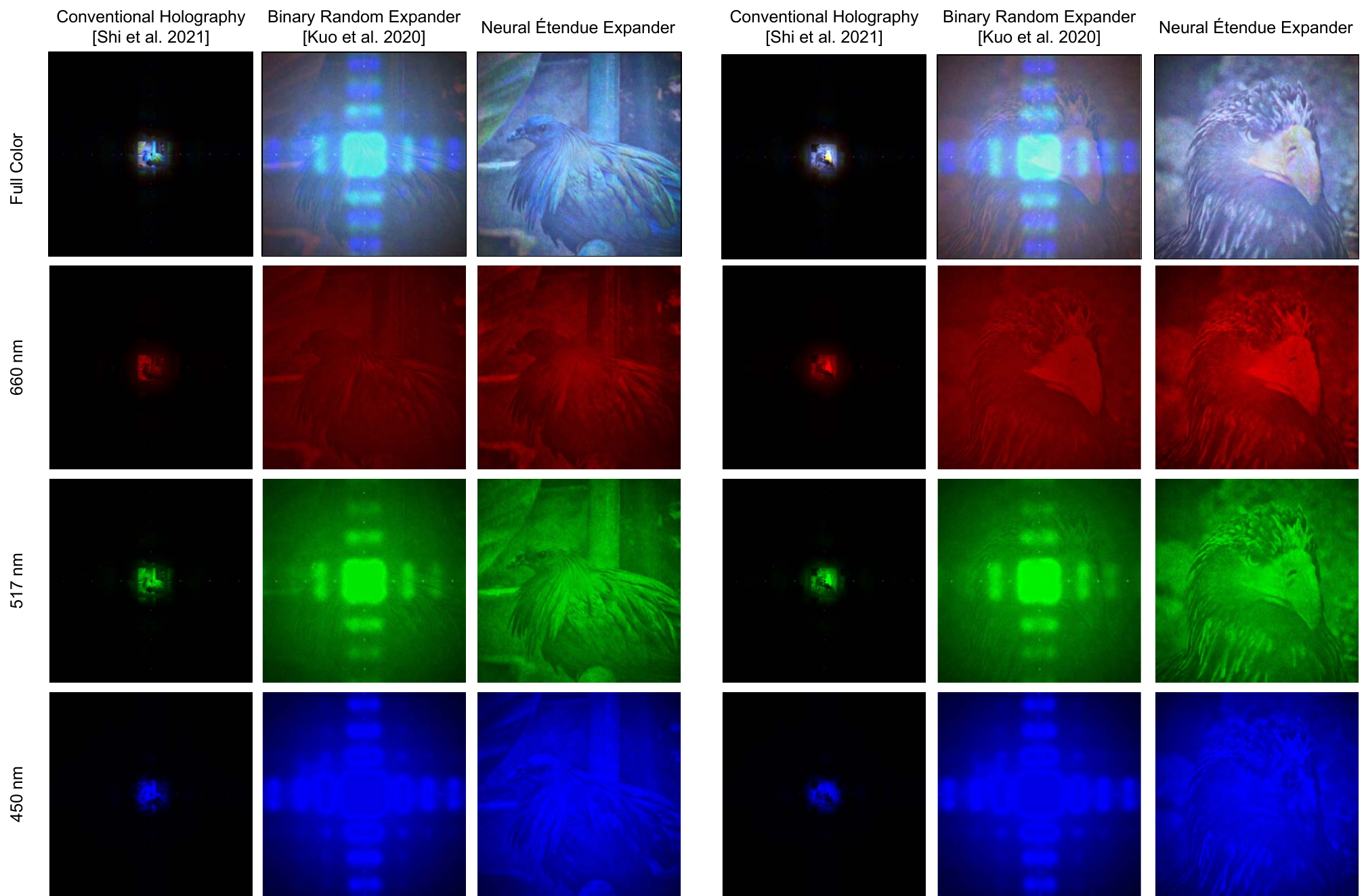}
    \fi
    \caption{\label{fig:real_color_64_decomp}
Experimentally captured color holograms at $64\times$ \'{e}tendue expansion. This figure shows the reconstructions for individual color channels for the holograms shown in Supplementary Fig.~\ref{fig:real_color_64}. These results supplement the experimental findings from Fig.~2 of the main manuscript.
}		
\end{figure*}

\begin{figure*}
	\centering
    	\if\loadFigures0
    \else
    \includegraphics[width=\linewidth]{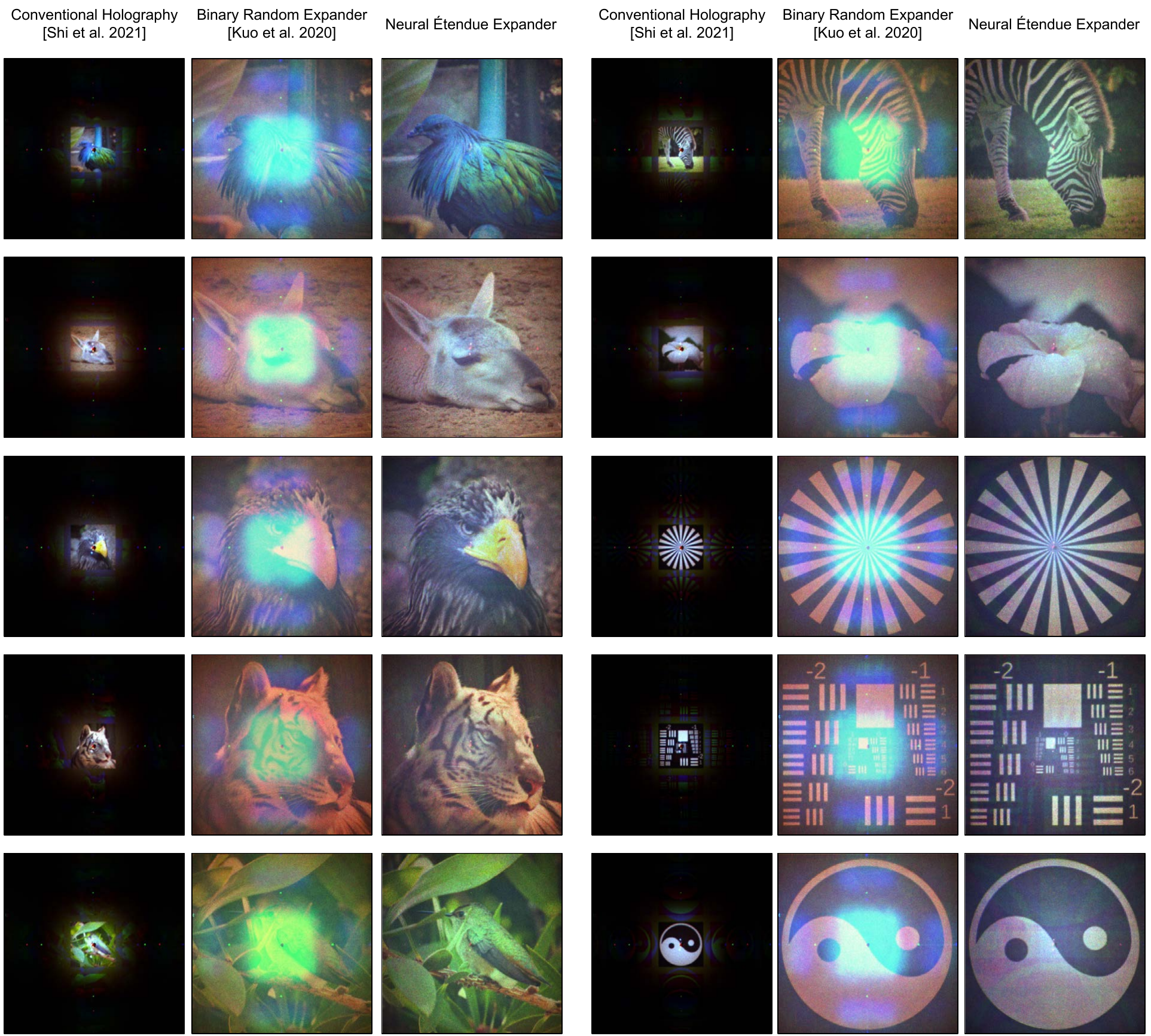}
    \fi
    \caption{\label{fig:real_color_16}
Experimentally captured color holograms at $16\times$ \'{e}tendue expansion. The wavelengths used are $\SI{660}{nm}$, $\SI{517}{nm}$, and $\SI{450}{nm}$. For comparison, \'{e}tendue expanded holograms generated with random expanders\cite{kuo2020expansion} and non-\'{e}tendue expanded holograms\cite{shi2021towards} are included. These results supplement the experimental findings from Fig.~2 of the main manuscript.
}		
\end{figure*}

\begin{figure*}
	\centering
    	\if\loadFigures0
    \else
    \includegraphics[width=\linewidth]{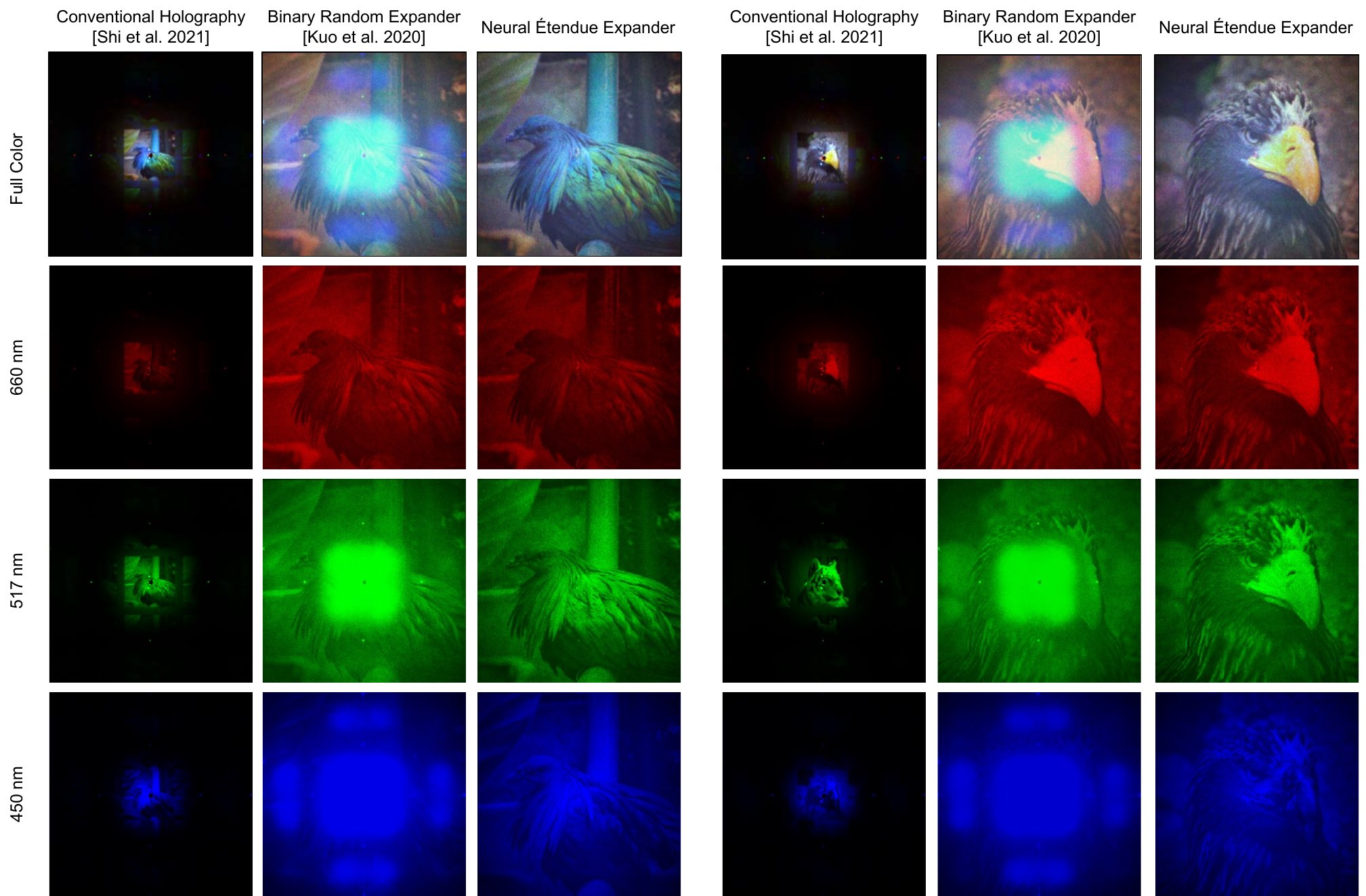}
    \fi
    \caption{\label{fig:real_color_16_decomp}
Experimentally captured color holograms at $16\times$ \'{e}tendue expansion. This figure shows the reconstructions for individual color channels for the holograms shown in Supplementary Fig.~\ref{fig:real_color_16}. These results supplement the experimental findings from Fig.~2 of the main manuscript. 
}		
\end{figure*}

\begin{figure*}
	\centering
    	\if\loadFigures0
    \else
    \includegraphics[width=\linewidth]{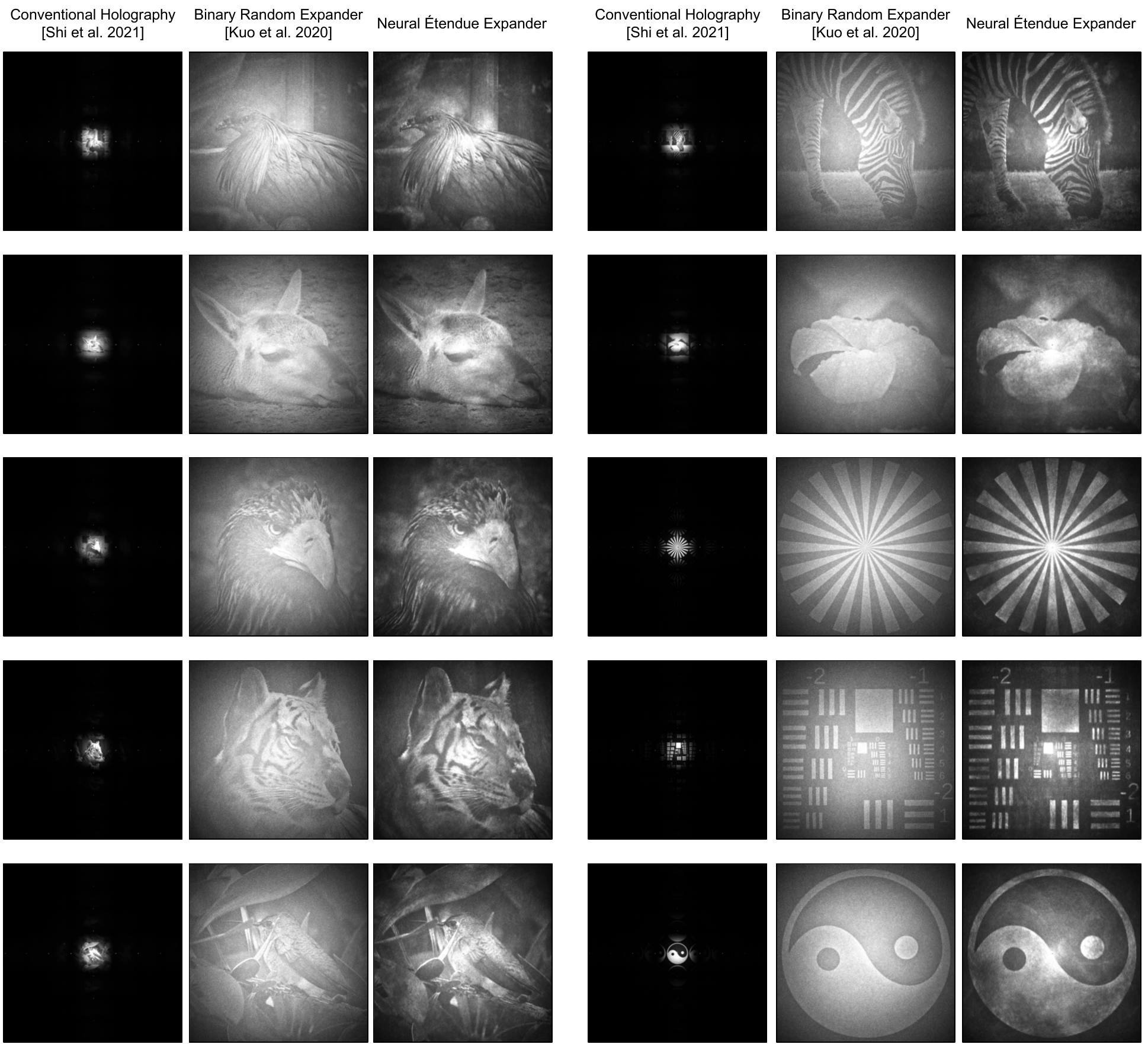}
    \fi
    \caption{\label{fig:real_mono_64}
Experimentally captured monochromatic holograms at $64\times$ \'{e}tendue expansion. The wavelengths used is $\SI{660}{nm}$. For comparison, \'{e}tendue expanded holograms generated with random expanders\cite{kuo2020expansion} and non-\'{e}tendue expanded holograms\cite{shi2021towards} are included. These results supplement the experimental findings from Fig.~2 of the main manuscript.
}		
\end{figure*}

\begin{figure*}
	\centering
    	\if\loadFigures0
    \else
    \includegraphics[width=\linewidth]{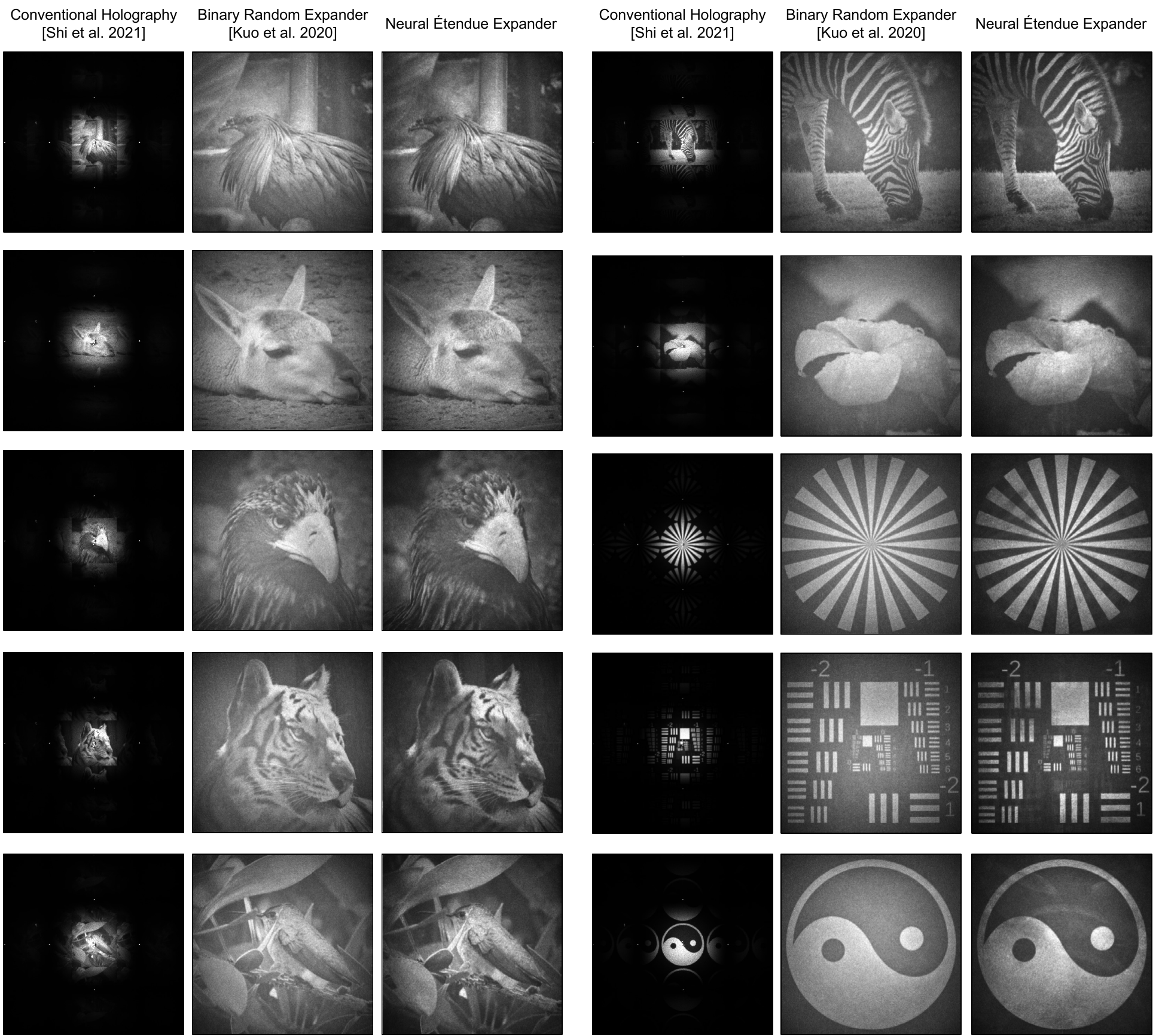}
    \fi
    \caption{\label{fig:real_mono_16}
Experimentally captured monochromatic holograms at $16\times$ \'{e}tendue expansion. The wavelengths used is $\SI{660}{nm}$. For comparison, \'{e}tendue expanded holograms generated with random expanders\cite{kuo2020expansion} and non-\'{e}tendue expanded holograms\cite{shi2021towards} are included. These results supplement the experimental findings from Fig.~2 of the main manuscript.
}		
\end{figure*}

Next, we provide addition experimental findings that were acquired on the experimental setup. Supplementary Fig.~\ref{fig:real_color_64} reports $64\times$ \'{e}tendue expanded color holograms. Supplementary Fig.~\ref{fig:real_color_64_decomp} lists the reconstructions for individual color channels. Supplementary Fig.~\ref{fig:real_color_16} and Fig.~\ref{fig:real_color_16_decomp} display the corresponding color holograms at $16\times$ \'{e}tendue expansion. Supplementary Fig.~\ref{fig:real_mono_64} and Fig.~\ref{fig:real_mono_16} show $64\times$ and $16\times$ \'{e}tendue expanded monochromatic holograms.

We use temporal multiplexing to reduce speckle for all results shown. Specifically, all experimental results shown are computed as the average of 20 holograms, where each hologram is generated with a different random seed. Using a different random seed for each hologram results in different speckle artifacts which are then eliminated through temporal multiplexing.

In addition, we performed white balancing on the captures so that the overall color ratio matches that of the target images. In the future, this color balancing step could have been performed in the hardware via laser power adjustment if we had a programmable laser source.

Lastly, all of the expanders leak a residual DC term similar to how the SLM also leaks a DC term. We suppressed this term by determining the hologram that corresponds to each expander's DC term and then subtracting it from the captured holograms. In the future, the DC term could be eliminated in hardware by using a tilted off-axis construction.

\clearpage

\section*{Supplementary Note 2: Additional Synthetic Results}
\label{sec:synthetic}
\begin{figure*}
	\centering
    	\if\loadFigures0
    \else
    \includegraphics[width=0.8\linewidth]{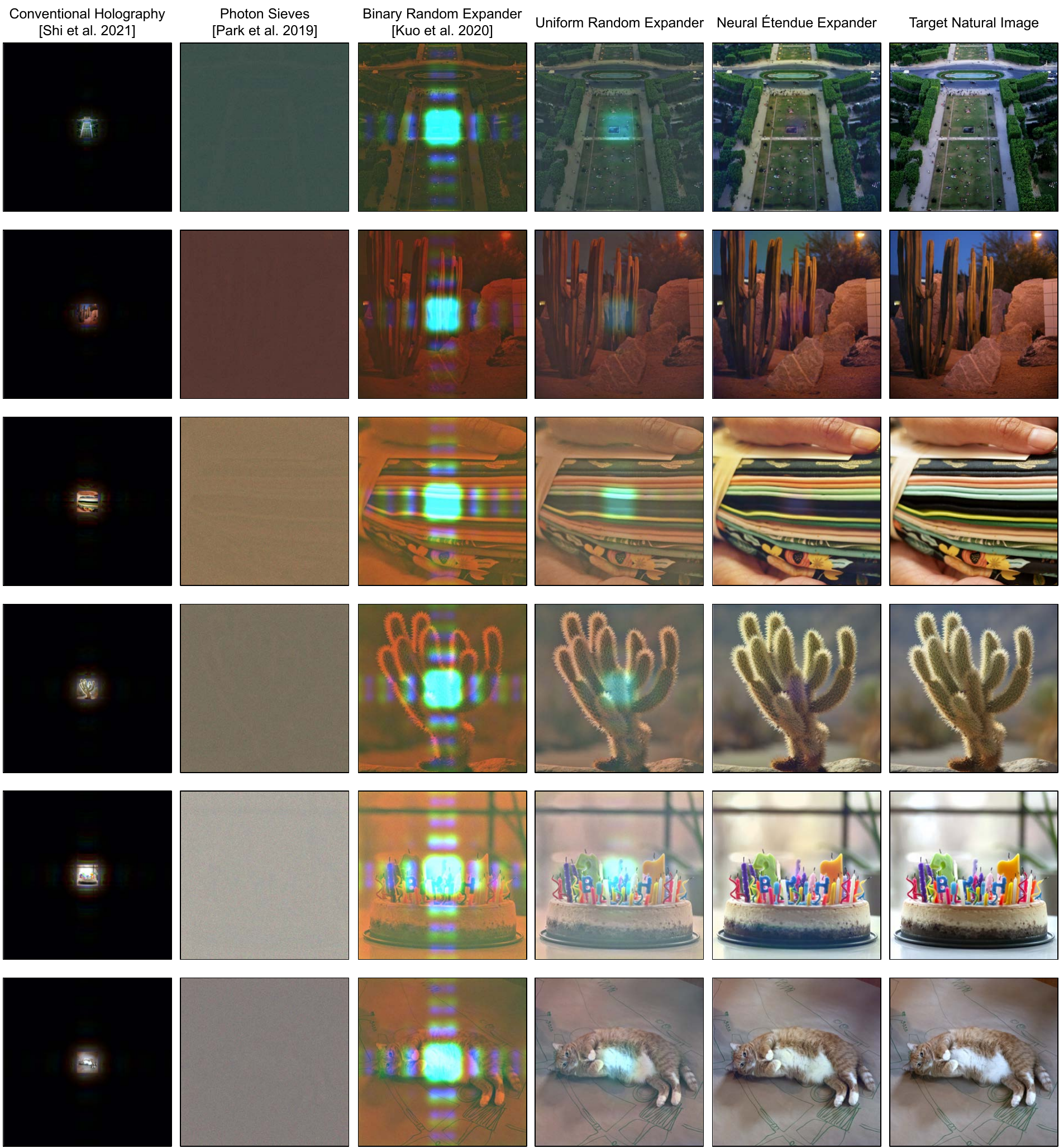}
    \fi
    \caption{\label{fig:sim_color_64}
Simulated color holograms at $64\times$ \'{e}tendue expansion. The wavelengths used are $\SI{660}{nm}$, $\SI{517}{nm}$, and $\SI{450}{nm}$. These experimental findings supplement the simulation results from Fig.~3 of the main manuscript.
}		
\end{figure*}

\begin{figure*}
	\centering
    	\if\loadFigures0
    \else
    \includegraphics[width=0.8\linewidth]{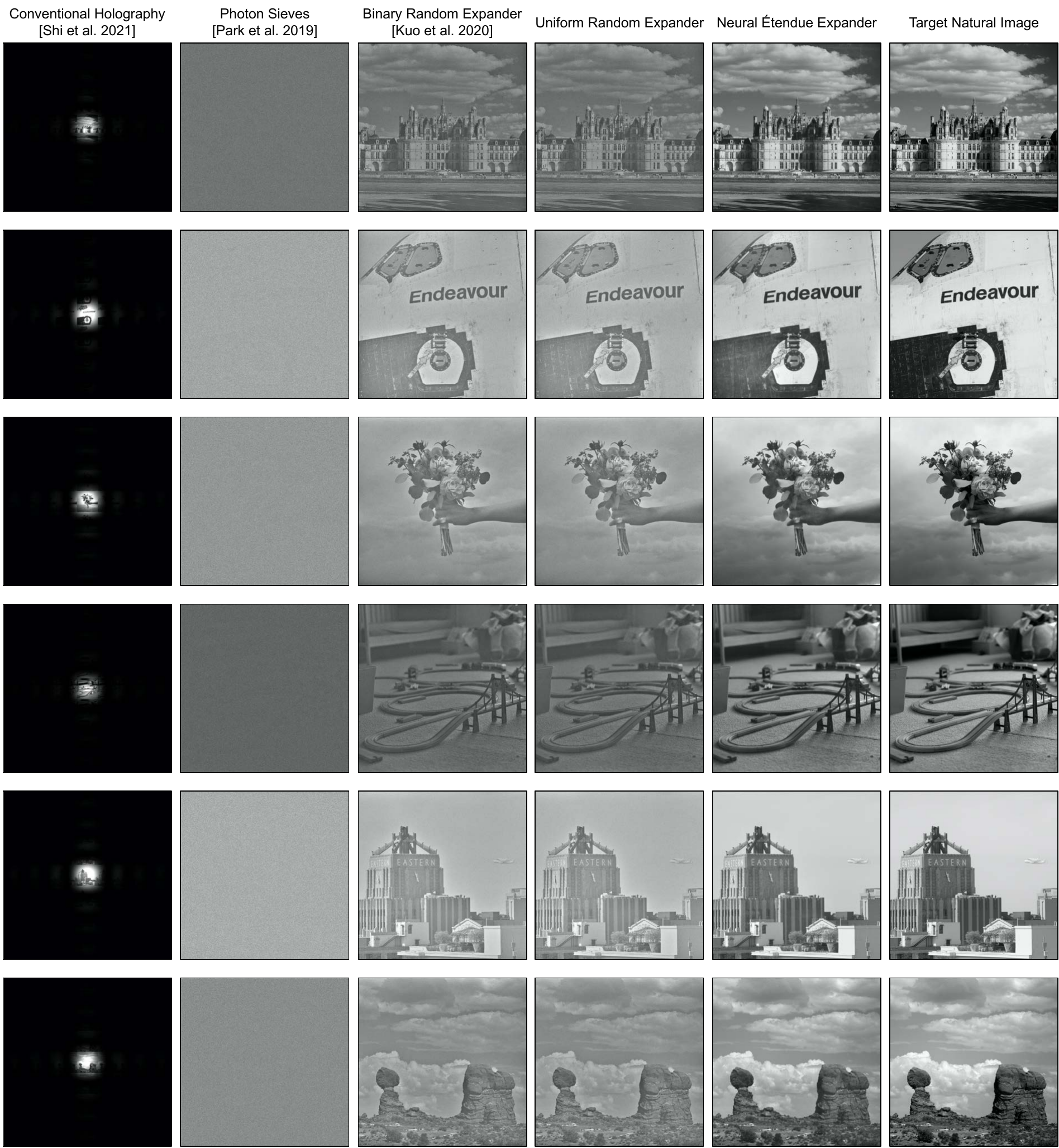}
    \fi
    \caption{\label{fig:sim_mono_64}
Simulated monochromatic holograms at $64\times$ \'{e}tendue expansion. The wavelength used is $\SI{660}{nm}$. These experiments supplement the simulation results from Fig.~3 of the main manuscript.
}		
\end{figure*}

\begin{figure*}
	\centering
    	\if\loadFigures0
    \else
    \includegraphics[width=0.8\linewidth]{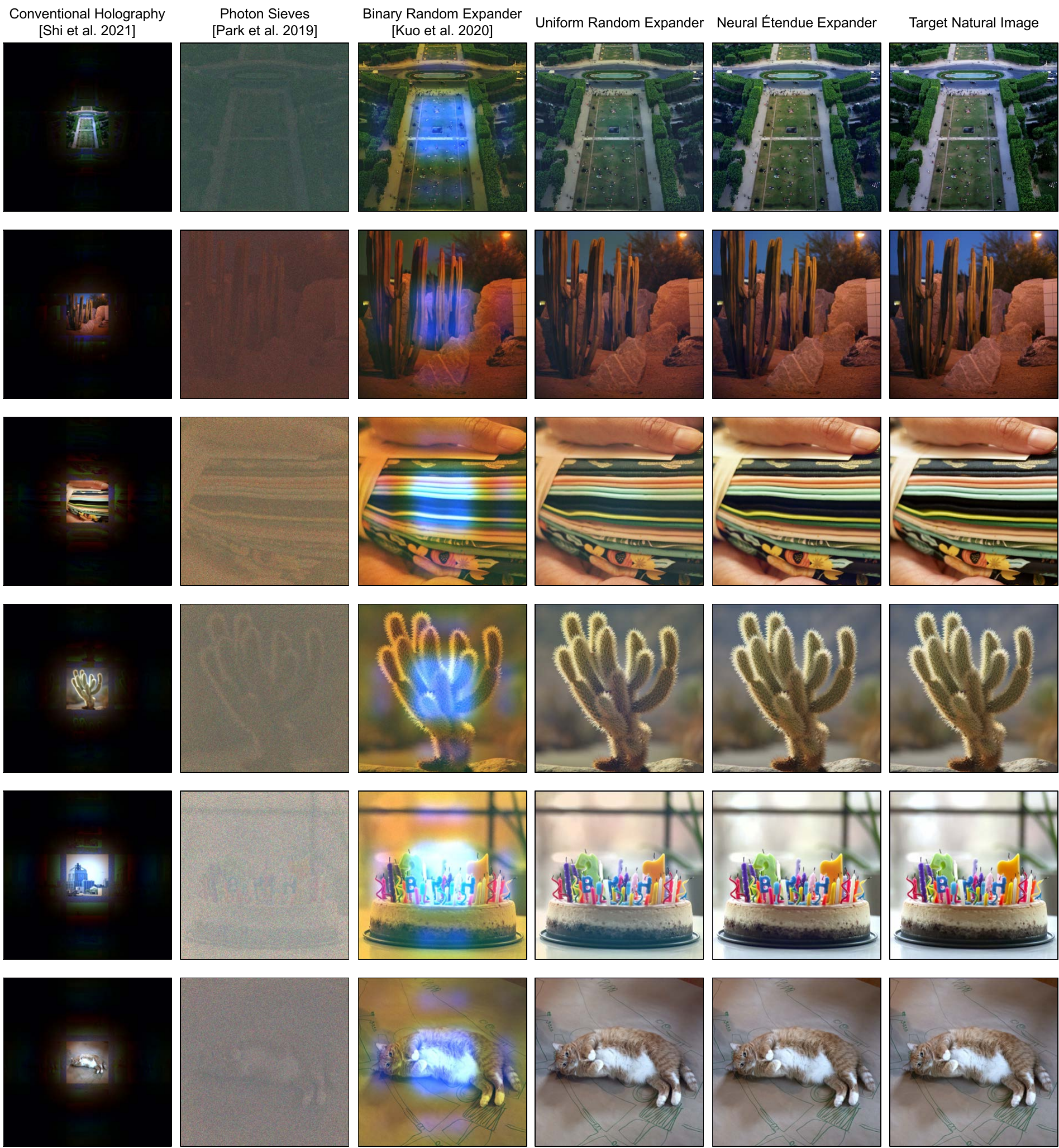}
    \fi
    \caption{\label{fig:sim_color_16}
Simulated color holograms at $16\times$ \'{e}tendue expansion. The wavelengths used are $\SI{660}{nm}$, $\SI{517}{nm}$, and $\SI{450}{nm}$. These experiments supplement the simulation results from Fig.~3 of the main manuscript.
}		
\end{figure*}

\begin{figure*}
	\centering
    	\if\loadFigures0
    \else
    \includegraphics[width=0.8\linewidth]{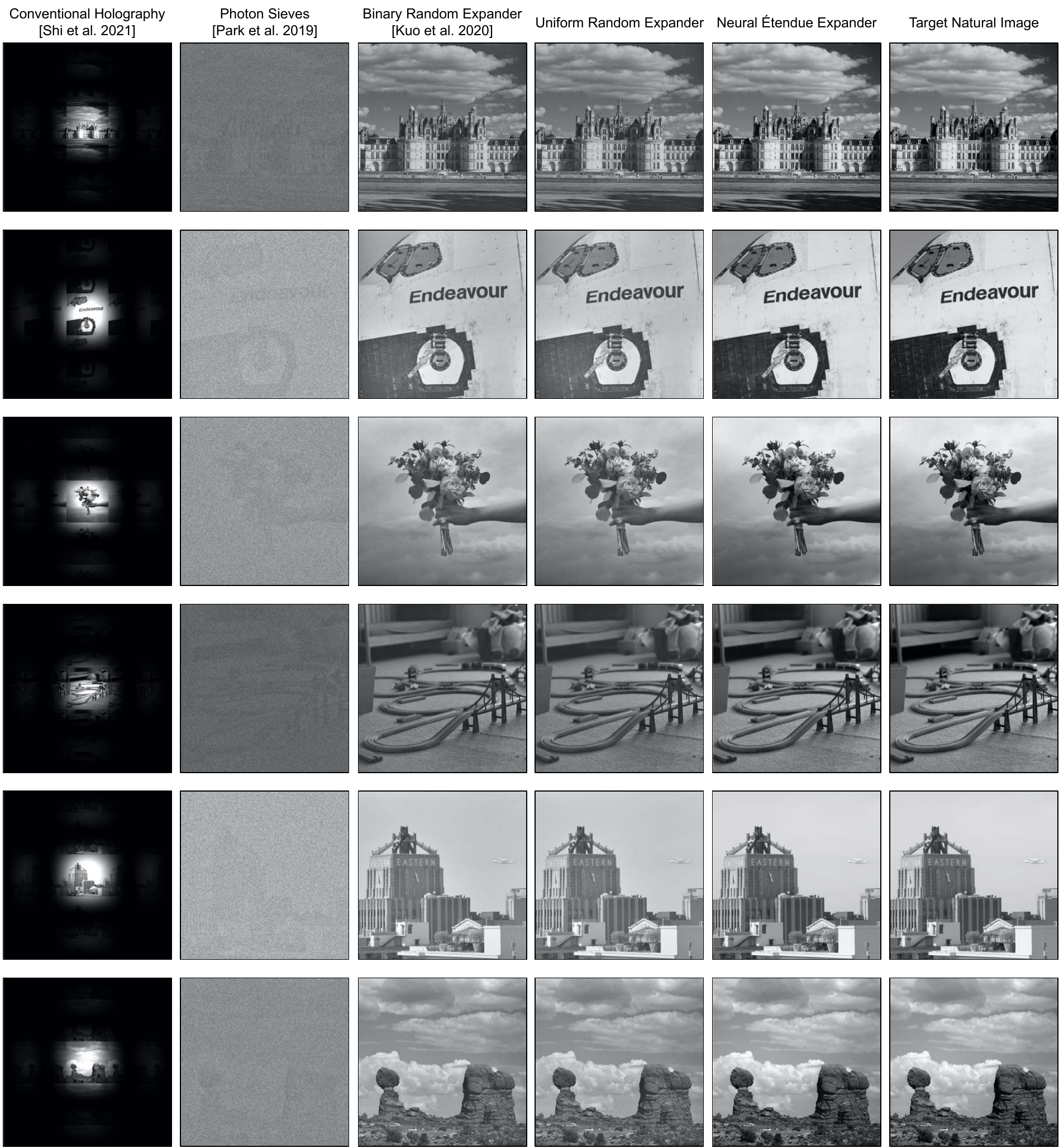}
    \fi
    \caption{\label{fig:sim_mono_16}
Simulated monochromatic holograms at $16\times$ \'{e}tendue expansion. The wavelength used is $\SI{660}{nm}$. These experiments supplement the simulation results from Fig.~3 of the main manuscript.
}		
\end{figure*}

We provide addition simulation results that further validate the effects of neural \'{e}tendue expansion. Supplementary Fig.~\ref{fig:sim_color_64} compares $64\times$ \'{e}tendue expansion with neural \'{e}tendue expanders against \'{e}tendue expansion with photon sieves\cite{Park2019UltrathinWL} and randomized expanders\cite{kuo2020expansion}. Supplementary Fig.~\ref{fig:sim_mono_64} reports the same comparison for only a single wavelength. Supplementary Figs.~\ref{fig:sim_color_16} and \ref{fig:sim_mono_16} report the same comparisons for $16\times$ \'{e}tendue expansion.

The wavelengths used are the same as in the physical experiment, specifically $\SI{660}{nm}$, $\SI{517}{nm}$, and $\SI{450}{nm}$. The random expanders here are also designed for $\SI{660}{nm}$ as in the physical experiment. We assume $100\%$ diffraction efficiency for both the SLM and the expander so no DC block is simulated.

Quantitative scores are shown in Supplementary Table~\ref{tab:64x_and_16x_quant}. The monochromatic evaluation uses a neural \'{e}tendue expander designed specifically for a single wavelength whereas the trichromatic evalutions use a neural \'{e}tendue expander designed simultaneously for all three wavelengths. The random expander\cite{kuo2020expansion} can only be designed for one wavelength and in this case it is designed for $\SI{660}{nm}$. The photon sieve\cite{Park2019UltrathinWL} is agnostic to wavelength because it only affects the amplitude component. The training and test datasets consists of images from personal photo collections, the DIV2K dataset\cite{Agustsson_2017_CVPR_Workshops}, and the INRIA Holiday dataset\cite{Jgou2008HammingEA}.

The CGH algorithm used for our method is end-to-end gradient descent optimization with our fully differentiable image formation model\cite{chakravarthula2019wirtinger}. We apply the same CGH algorithm when running baseline comparisons for the random expander. For the photon sieve, we follow the CGH algorithm described in their Supplementary Information. Specifically, we compute a phase-only inverse Fourier transform of the target image using IFTA\cite{Wyrowski_1988_ifta} and then we directly sample the Fourier spectrum at the locations of the holes in the photon sieve.

We incorporate wavelength dependent dispersion into these simulations. As such, the full color holograms are generated by slightly reducing the FOV of the reconstructed holograms of the red channel ($\SI{660}{nm}$) than the green channel ($\SI{520}{nm}$). Specifically, the FOV for the red channel is scaled by a factor of $\SI{450}{nm} / \SI{660}{nm}$ and the FOV for the green channel is scaled by a factor of $\SI{520}{nm} / \SI{660}{nm}$. For the monochromatic evaluation the full FOV of the red channel ($\SI{660}{nm}$) is used. 

\begin{table}
\small
\caption{Quantitative evaluation of $64\times$ and $16\times$ \'{e}tendue expansion reconstructions. All reconstructions are evaluated after frequency filtering. Error bars correspond to one standard deviation. The highest performing score for each setting is highlighted in \textbf{bold}.}
\vspace{8pt}
\label{tab:64x_and_16x_quant}
\scriptsize
\centering
\renewcommand{\arraystretch}{1.5}
  \begin{tabular}{ccccc}
  \toprule
                                              & \makecell[c]{Neural \'{E}tendue\\Expander} & \makecell[c]{Uniform Random\\Expander} & \makecell[c]{Binary Random\\Expander\cite{kuo2020expansion}} & Photon Sieves\cite{Park2019UltrathinWL} \\ \midrule
  \makecell[c]{$64\times$ Monochromatic\\$\SI{660}{nm}$} & $\bf{29.1} \pm 2.3$ & $17.4 \pm 1.6$ & $17.4 \pm 1.6$ & $13.5 \pm 1.1$ \\ [10pt]
  \makecell[c]{$64\times$ Trichromatic\\All Colors}      & $\bf{31.6} \pm 3.6$ & $16.8 \pm 1.5$ & $13.6 \pm 1.2$ & $13.1 \pm 1.0$ \\ [10pt]
  \makecell[c]{$64\times$ Trichromatic\\$\SI{660}{nm}$}  & $\bf{39.9} \pm 8.4$ & $19.9 \pm 2.9$ & $19.8 \pm 2.9$ & $13.0 \pm 1.5$ \\ [10pt]
  \makecell[c]{$64\times$ Trichromatic\\$\SI{520}{nm}$}  & $\bf{33.4} \pm 4.2$ & $16.0 \pm 1.6$ & $11.9 \pm 0.9$ & $13.4 \pm 1.1$ \\ [10pt]
  \makecell[c]{$64\times$ Trichromatic\\$\SI{450}{nm}$}  & $\bf{21.6} \pm 4.4$ & $14.6 \pm 2.2$ & $9.2  \pm 1.7$  & $12.8 \pm 1.8$ \\ \midrule
  \makecell[c]{$16\times$ Monochromatic\\$\SI{660}{nm}$} & $\bf{49.7} \pm 4.0$ & $25.6 \pm 3.8$ & $25.5 \pm 3.8$ & $13.5 \pm 1.0$ \\ [10pt]
  \makecell[c]{$16\times$ Trichromatic\\All Colors}      & $\bf{47.4} \pm 4.2$ & $31.9 \pm 5.0$ & $24.2 \pm 3.4$ & $13.2 \pm 1.0$ \\ [10pt]
  \makecell[c]{$16\times$ Trichromatic\\$\SI{660}{nm}$}  & $\bf{54.8} \pm 2.9$ & $47.3 \pm 9.6$ & $46.8 \pm 9.9$ & $13.3 \pm 1.4$ \\ [10pt]
  \makecell[c]{$16\times$ Trichromatic\\$\SI{520}{nm}$}  & $\bf{51.9} \pm 4.0$ & $27.8 \pm 6.3$ & $16.1 \pm 1.8$ & $13.5 \pm 1.1$ \\ [10pt]
  \makecell[c]{$16\times$ Trichromatic\\$\SI{450}{nm}$}  & $\bf{35.6} \pm 9.2$ & $20.6 \pm 4.3$ & $9.8  \pm 1.7$ & $12.8 \pm 1.8$ \\
  \bottomrule  \end{tabular}
\end{table}

\clearpage

\section*{Supplementary Note 3: Expander Analysis}
\label{sec:analysis}
In this section we provide additional analysis and derivations of the virtual frequency upper bound and the display specifications when our method is integrated with an 8K SLM\cite{buckley2006viewingAngle}.

\paragraph{Derivation of Virtual Frequency Upper Bound} Here we provide further steps to show how we obtained the upper bound described by Equation 2 in the main manuscript.
\begin{align}\label{eq:proof_deriv}
 \mathcal{L}_T &= \min \limits_{\mathcal{S}} \left\| {\left( {|{\mathcal{F}}\left( {{\mathcal{ E}} \odot U\left( {{\mathcal{S}}} \right)} \right){|^2} - {T}} \right)*f} \right\|_2^2 \nonumber \\
  &\leq \left\| {\left( {|{\mathcal{F}}\left( \mathcal{E} \right){|^2} - {T}} \right)*f} \right\|_2^2 \nonumber \\
  &= \frac{1}{N}\left\| {\left( {\mathcal{F}(|{\mathcal{F}}\left( \mathcal{E} \right){|^2}) - \mathcal{F}(T)} \right) \odot \mathcal{F}(f)} \right\|_2^2 \nonumber \\
  &= \frac{1}{N}\left\| {\left( {\widetilde{\mathcal{E}} - \mathcal{F}(T)} \right) \odot \mathcal{F}(f)} \right\|_2^2.
\end{align}
The inequality is obtained because the loss value that is obtained with the optimum setting for $\mathcal{S}$ must be less than or equal to the loss value when $\mathcal{S}$ is uses the zero-phase setting. This upper bound manifests itself in the learned frequency spectrum of the neural \'{e}tendue expanders. See Supplementary Fig.~\ref{fig:virtural_freq} for a comparison between the virtual frequency modulation of the learned neural \'{e}tendue expanders and the virtual frequency modulation of natural images.

\paragraph{\'{E}tendue-expanded Display Specifications} Here we provide the derivation of the FOV and eyebox sizes after \'{e}tendue expansion. We first assume an 8K SLM\cite{buckley2006viewingAngle} with $7680\times4320$ pixel resolution at $\SI{660}{nm}$ wavelength. We augment the SLM with a $64\times$ \'{e}tendue expander, resulting in $61440\times34560$ pixel resolution. Now the FOV and eyebox size are related to each other by
\begin{equation}
\text{FOV} \times \text{eyebox} = \lambda \times N_n,
\end{equation}
where $N_n$ is the expanded pixel resolution\cite{kuo2020expansion}. By setting the eyebox to $\SI{18.5}{mm}$ we get a horizontal FOV of $126^\circ$ and a vertical FOV of $71^\circ$. The stereo FOV provided is given by
\begin{equation}
\text{Stereo FOV} = 4 \arcsin(\sin(\text{Horizontal FOV} / 2) \times \sin(\text{Vertical FOV} / 2)).
\end{equation}
Thus, the total stereo FOV provided is 2.175 steradians. Human stereo FOV is 2.56 steradians\cite{nasa2011space} so this \'{e}tendue expanded holographic display would provide a stereo FOV that covers $85\%$ of the human stereo FOV. Note that this value could vary per individual.

For this same system, the provided angular resolution is given by
\begin{equation}
\text{Angular Resolution} = N_s / \text{FOV}
\end{equation}
where $N_s$ is the native SLM resolution. Thus, the angular resolution is 61 pixels/degree for the setup assumed above.

\clearpage

\begin{figure*}
	\centering
    	\if\loadFigures0
    \else
    \includegraphics[width=\linewidth]{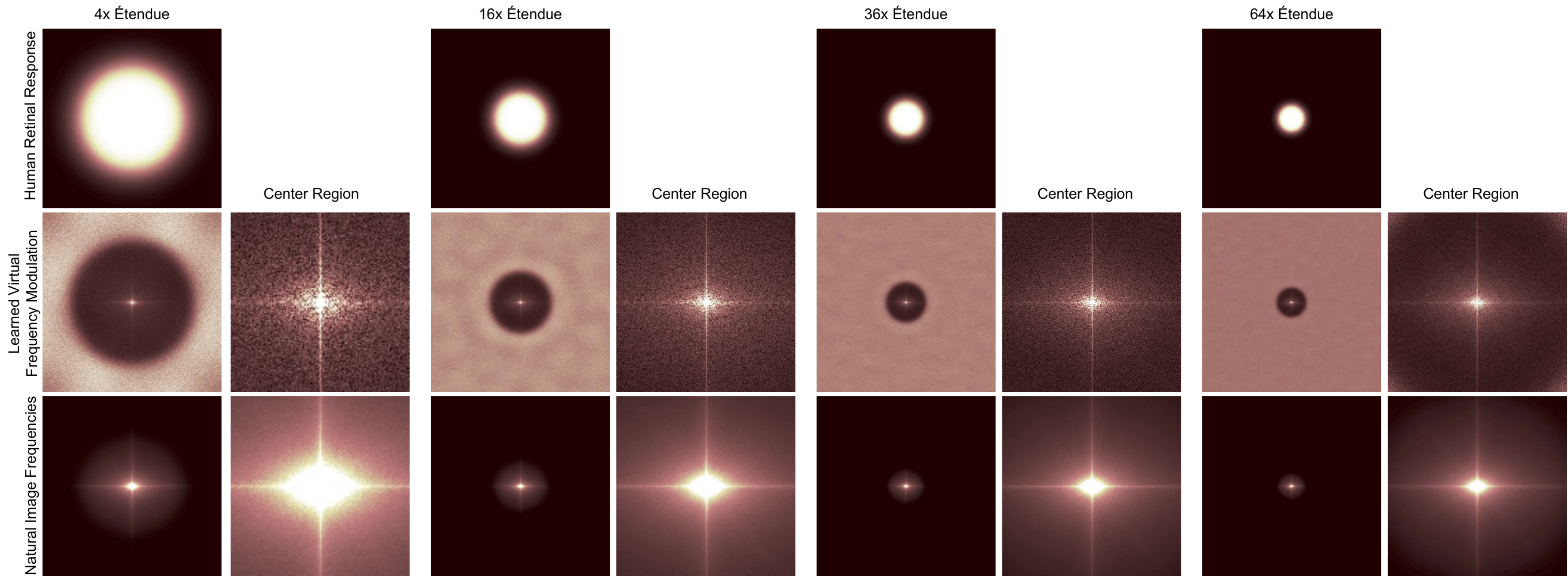}
    \fi
	\caption{\label{fig:virtural_freq}
	Comparison of virtual frequency modulation of neural \'{e}tendue expander and natural images. Most content for natural images resides within the lower frequency bands. The human visual system is also largely biased towards the lower frequency terms. We coarsely approximate the human visual response with low pass filters (top row). Our training algorithm teaches the expander to learn a virtual frequency modulation (middle row) that approximates the frequency modulation of a natural image dataset (bottom row). The center region insets correspond to the center eighth of each spectrum.
}		
\end{figure*}

\clearpage

\section*{Supplementary Note 4: \'{E}tendue Expansion for 3D Holograms}
\label{sec:volume}
Neural \'{e}tendue expanders also facilitate the generation of 3D holograms. To demonstrate this, we optimize the SLM to produce 3D focal stack holograms.
Specifically, we solve a variant of Eq.~(3) from the main manuscript. This variant is given by
\begin{equation}\label{eq:opt_3D}
\mathop \mathrm{minimize} \limits_{\mathcal{E},\mathcal{S}} {\left\| {\left( {|{\mathcal{F}}\left( {{\mathcal{ E}} \odot U\left( {{\mathcal{S}}} \right)} \right){|^2} - {T_\mathrm{near}}} \right)*f} \right\|_2^2} + {\left\| {\left( {|{\mathcal{F}}\left( {\mathcal{Z} \odot {\mathcal{ E}} \odot U\left( {{\mathcal{S}}} \right)} \right){|^2} - {T_\mathrm{far}}} \right)*f} \right\|_2^2},
\end{equation}
where $T_\mathrm{near}$ is the target image at the near plane, $T_\mathrm{far}$ is the target image at the far plane, and $Z$ is a z-offset phase given by
\begin{equation}\label{eq:z_phase}
\mathcal{Z}(x,y) = \frac{2\pi}{\lambda} \frac{\left( x^2 + y^2\right)}{2f}
\end{equation}
where $f$ is the z-distance of the defocus and $\lambda$ is the wavelength of the laser source. Within the HOLOEYE software the distance $f$ can be set with a slider $s$ as
\begin{equation}\label{eq:focal_length}
f = \frac{\max(h,w) \cdot \Delta_\mathrm{SLM}}{\lambda \cdot s}
\end{equation}
where $h$ is the height of the SLM, $w$ is the width of the SLM, and $\Delta_\mathrm{SLM}$ is the SLM pixel pitch. Solving Equation~\ref{eq:opt_3D} will produce a two plane \'{e}tendue expanded 3D hologram. Simulated qualitative results are shown in Supplementary Fig.~\ref{fig:sim_3D}. We used $s = -0.005$ for these results. The physical separation of the two focal planes also depends on the focal length of the Fourier Transforming lens. For the $\SI{80}{mm}$ lens used in our setup this setting of $s$ results in a physical separation of approximately $\SI{3}{mm}$.

Neural \'{e}tendue expansion achieves high-fidelity \'{e}tendue expanded 3D color holograms. We measure over 5 dB improvement in PSNR over 3D color holograms generated with uniform random expanders. We note that no existing \'{e}tendue expansion technique has successfully demonstrated 3D color holography aside from photon sieves. However, the photon sieve expansion method only supports 3D holography of sparse points and not natural scenes.

\begin{figure*}
	\centering
    	\if\loadFigures0
    \else
    \includegraphics[width=\linewidth]{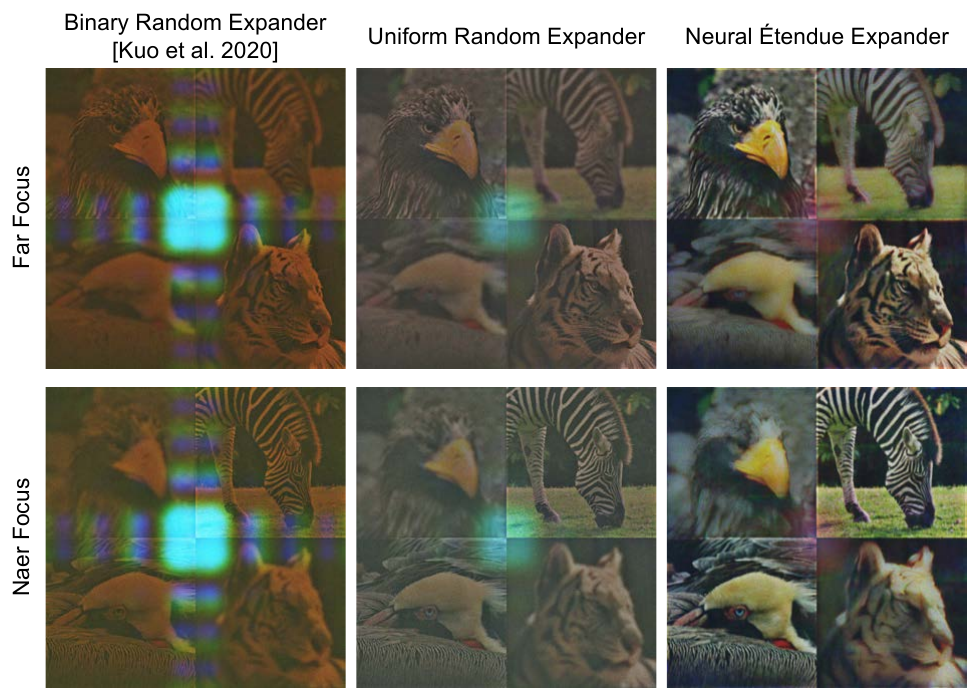}
    \fi
	\caption{\label{fig:sim_3D}
	Simulated results for \'{e}tendue expanded 3D color holograms. Here we show trichromatic holograms generated at $\SI{660}{nm}$, $\SI{517}{nm}$, and $\SI{450}{nm}$ at two different depth planes. The near plane corresponds to $\SI{0}{mm}$ and the far plane corresponds to a separation of $\SI{3}{mm}$ when using an $\SI{80}{mm}$ focal length lens as the Fourier Transforming lens. Similar to the \'{e}tendue expanded 2D hologram results, neural \'{e}tendue expanders are capable of improving the fidelity of \'{e}tendue expanded 3D holograms of natural scenes. In contrast, the uniform and binary random expanders achieve lower fidelity.
}		
\end{figure*}

\clearpage

\section*{Supplementary Note 5: Eyebox Analysis}
\label{sec:eyebox}
In this section we evaluate the robustness of different \'{e}tendue expansion methods to changes in the eye pupil position and size. We vary the eye pupil setting and we apply pupil-aware holography\cite{Chakravarthula2022PupilAwareH} to the holograms generated with binary random expanders\cite{kuo2020expansion}, quadratic phase profiles\cite{monin2022analyzing}, and neural \'{e}tendue expanders. Quantitative performance is shown in Supplementary Table~\ref{tab:16x_eyebox} and qualitative results are shown in Supplementary Figures~\ref{fig:sim_eyebox_1} and \ref{fig:sim_eyebox_2}.

We observe that neural \'{e}tendue expanders supports high-fidelity holograms varying pupil positions and sizes. This is because we initialized the learning process with a random expander which biases the final solution towards solutions that spread energy evenly throughout the eyebox. Quadratic phase profiles do not truly expand \'{e}tendue, they increase the FOV but change the energy distribution within the eyebox. Hence, changing the eye pupil setting causes image content to disappear, even when combined with pupil-aware holography.

\begin{figure*}
	\centering
    	\if\loadFigures0
    \else
    \includegraphics[width=\linewidth]{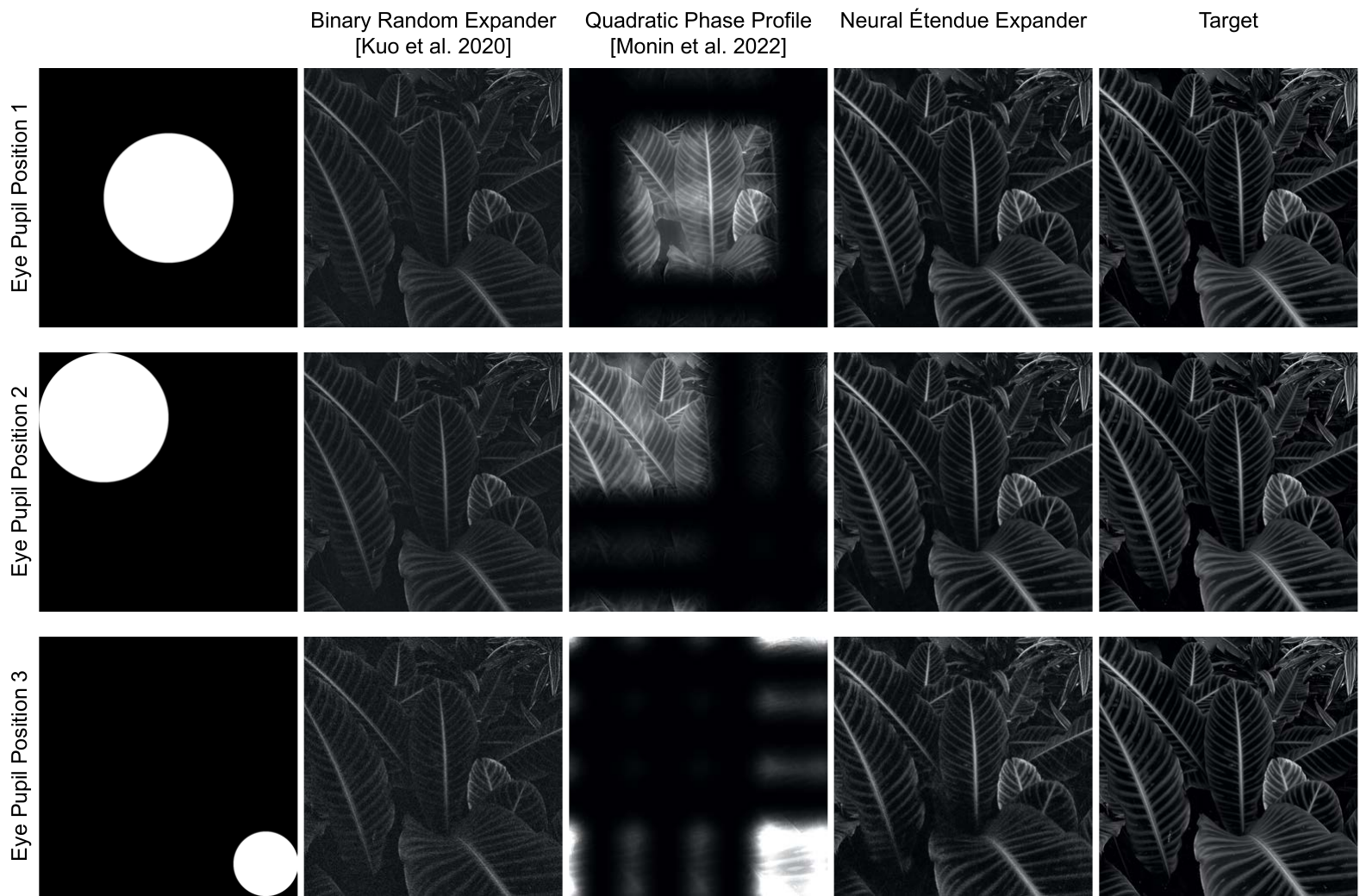}
    \fi
	\caption{\label{fig:sim_eyebox_1}
	Impact of eye pupil movement on perceived hologram. The eye pupil position and size is shown in the left column. The reconstructed \'{e}tendue expanded holograms are shown for binary random expanders\cite{kuo2020expansion}, quadratic phase profiles\cite{monin2022analyzing}, and neural \'{e}tendue expanders. Changes in the eye pupil position causes image content to disappear when using quadratic phase profiles. Neural \'{e}tendue expansion and random expansion both support a complete eyebox where the position of the eye pupil does not significantly affect the perceived image content. Furthermore, neural \'{e}tendue expansion outperforms random expansion at all eye pupil positions. These holograms are simulated at 16$\times$ \'{e}tendue expansion.
}
\end{figure*}

\begin{figure*}
	\centering
    	\if\loadFigures0
    \else
    \includegraphics[width=\linewidth]{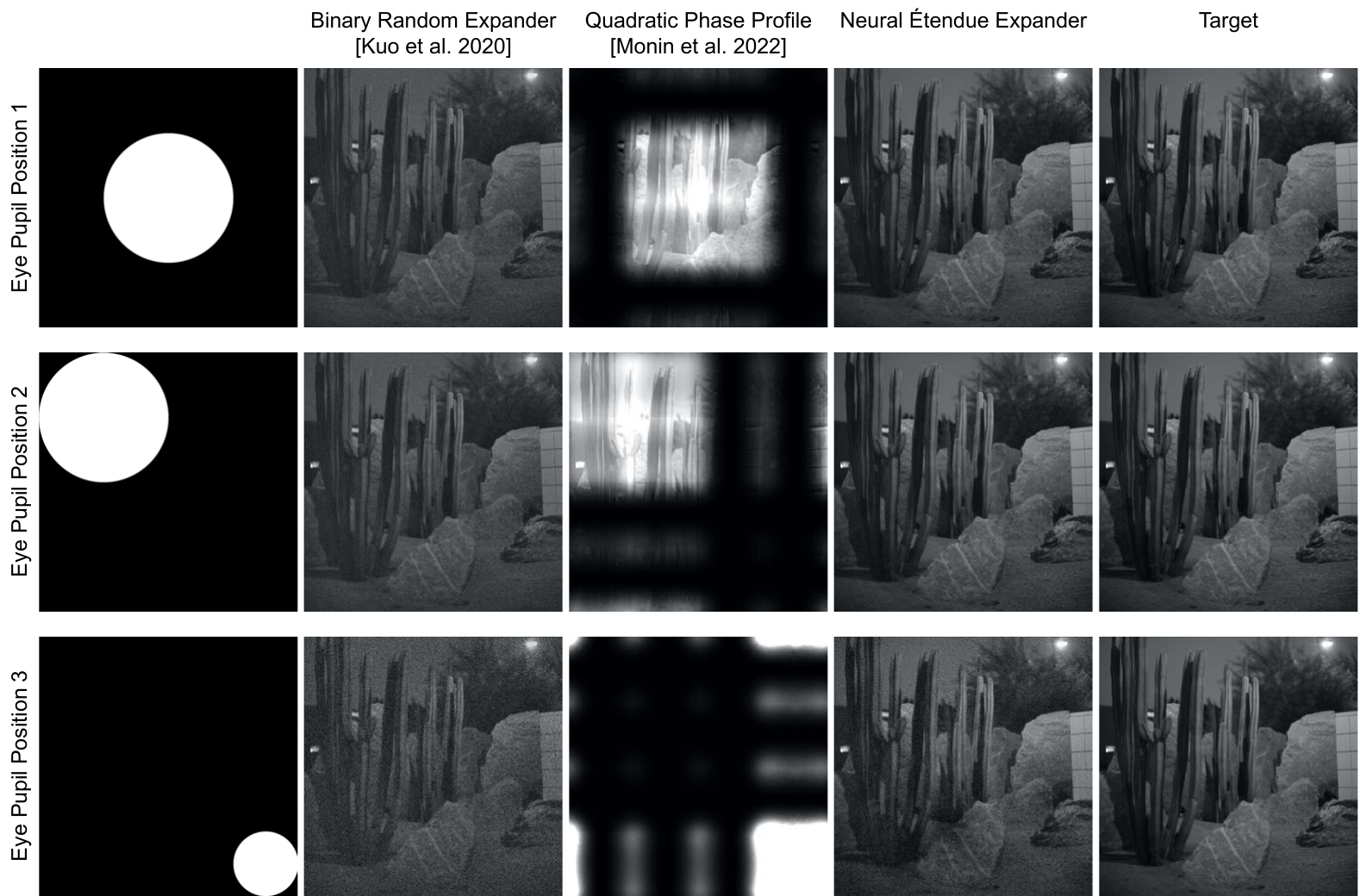}
    \fi
	\caption{\label{fig:sim_eyebox_2}
	Impact of eye pupil movement on perceived hologram. The eye pupil position and size is shown in the left column. The reconstructed \'{e}tendue expanded holograms are shown for binary random expanders\cite{kuo2020expansion}, quadratic phase profiles\cite{monin2022analyzing}, and neural \'{e}tendue expanders. Changes in the eye pupil position causes image content to disappear when using quadratic phase profiles. Neural \'{e}tendue expansion and random expansion both support a complete eyebox where the position of the eye pupil does not significantly affect the perceived image content. Furthermore, neural \'{e}tendue expansion outperforms random expansion at all eye pupil positions. These holograms are simulated at 16$\times$ \'{e}tendue expansion.
}
\end{figure*}

\begin{table}
\small
\caption{Quantitative evaluation of $16\times$ \'{e}tendue expansion reconstructions with different eye pupil positions. All reconstructions are evaluated after frequency filtering. Error bars correspond to one standard deviation. The highest performing score for each eye pupil setting is highlighted in \textbf{bold}.}
\vspace{8pt}
\label{tab:16x_eyebox}
\scriptsize
\centering
\renewcommand{\arraystretch}{1.5}
  \begin{tabular}{cccc}
  \toprule
                                              & \makecell[c]{Neural \'{E}tendue\\Expander} & \makecell[c]{Quadratic Phase\\Profile\cite{monin2022analyzing}} & \makecell[c]{Binary Random\\Expander\cite{kuo2020expansion}} \\ \midrule
  \makecell[c]{Eye Pupil\\Position 1} & $\bf{24.5} \pm 4.4$ & $7.3 \pm 2.3$ & $18.7 \pm 3.5$ \\ [10pt]
  \makecell[c]{Eye Pupil\\Position 2} & $\bf{24.7} \pm 4.6$ & $7.9 \pm 2.5$ & $18.7 \pm 3.5$ \\ [10pt]
  \makecell[c]{Eye Pupil\\Position 3} & $\bf{17.7} \pm 2.7$ & $7.1 \pm 1.7$ & $15.5 \pm 2.4$ \\ [10pt]
  \bottomrule  \end{tabular}
\end{table}

\clearpage

\section*{Supplementary Note 6: Resolution Analysis}
\label{sec:resolution}
In this section we simulate \'{e}tendue expansion with different native SLM resolutions from 1K-pixels to 8K-pixels. The results are reported in Supplementary Table~\ref{tab:slm_resolution}. We observe negligible changes in reconstruction fidelity across different SLM resolutions. This validates that the reconstruction fidelity
depends only on the \'{e}tendue expansion factor and not on the pixel resolution. Furthermore, this means that our technique when combined with an 8K SLM could provide the necessary stereo FOV and eyebox size needed for an immersive AR/VR experience, as per the specifications discussed in Supplementary Note 3.

While the perceived display quality of a practical system depends on many factors such as the SLM's pixel fill factor and the expander's diffraction efficiency, these are factors separate from the analysis that we are performing. Here, we assume ideal devices and hardware construction to demonstrate that the theoretical improvement in reconstruction fidelity for a given \'{e}tendue expansion factor does not change for different native SLM resolutions. Compensating for non-idealities in practical hardware can be accomplished with recent developments in neural holography\cite{chakravarthula2020learnedHIL,Peng:2020:NeuralHolography}.

\begin{table}
\small
\caption{Quantitative evaluation of $64\times$ \'{e}tendue expansion reconstructions for different native SLM resolutions. All reconstructions are evaluated after frequency filtering. Error bars correspond to one standard deviation. The highest performing score for each setting is highlighted in \textbf{bold}.}
\vspace{8pt}
\label{tab:slm_resolution}
\scriptsize
\centering
\renewcommand{\arraystretch}{1.5}
  \begin{tabular}{cccc}
  \toprule
   \makecell[c]{Native SLM\\Resolution}  & \makecell[c]{Neural \'{E}tendue\\Expander} & \makecell[c]{Uniform Random\\Expander} & \makecell[c]{Binary Random\\Expander\cite{kuo2020expansion}} \\ \midrule
  1K & $\bf{29.1} \pm 2.3$ & $17.4 \pm 1.6$ & $17.4 \pm 1.6$ \\ [10pt]
  2K & $\bf{29.1} \pm 2.3$ & $17.4 \pm 1.6$ & $17.4 \pm 1.5$ \\ [10pt]
  4K & $\bf{29.0} \pm 2.2$ & $17.5 \pm 1.5$ & $17.4 \pm 1.6$ \\ [10pt]
  8K & $\bf{29.0} \pm 2.2$ & $17.4 \pm 1.5$ & $17.4 \pm 1.6$ \\ [10pt]
  \bottomrule  \end{tabular}
\end{table}

\clearpage

\section*{Supplementary Note 7: \'{E}tendue Expansion with Higher Order Filtering}
\label{sec:echo}
The \'{e}tendue expanded holograms are formed as the convolution of the far field wavefronts of the expander and the SLM\cite{kuo2020expansion}. Both of these far fields contain repeated copies or echoes as shown in Supplementary Figure~\ref{fig:sim_echo}. When these far fields are convolved the echoes are intertwined, resulting in undesirable copies within the \'{e}tendue expanded hologram.

These copies do not noticeably degrade the quality of holograms of natural images. Nevertheless, if these copies must be removed then it can be performed by placing an amplitude block at the same location within the 4F system as the DC block. The amplitude block filters the higher order echoes coming from the SLM, as shown in Supplementary Figure~\ref{fig:sim_echo}. Simulated \'{e}tendue expanded holograms with this block are shown in Supplementary Figure~\ref{fig:sim_echo_results} and demonstrate that the undesired copies are removed.

\clearpage

\begin{figure*}
	\centering
    	\if\loadFigures0
    \else
    \includegraphics[width=\linewidth]{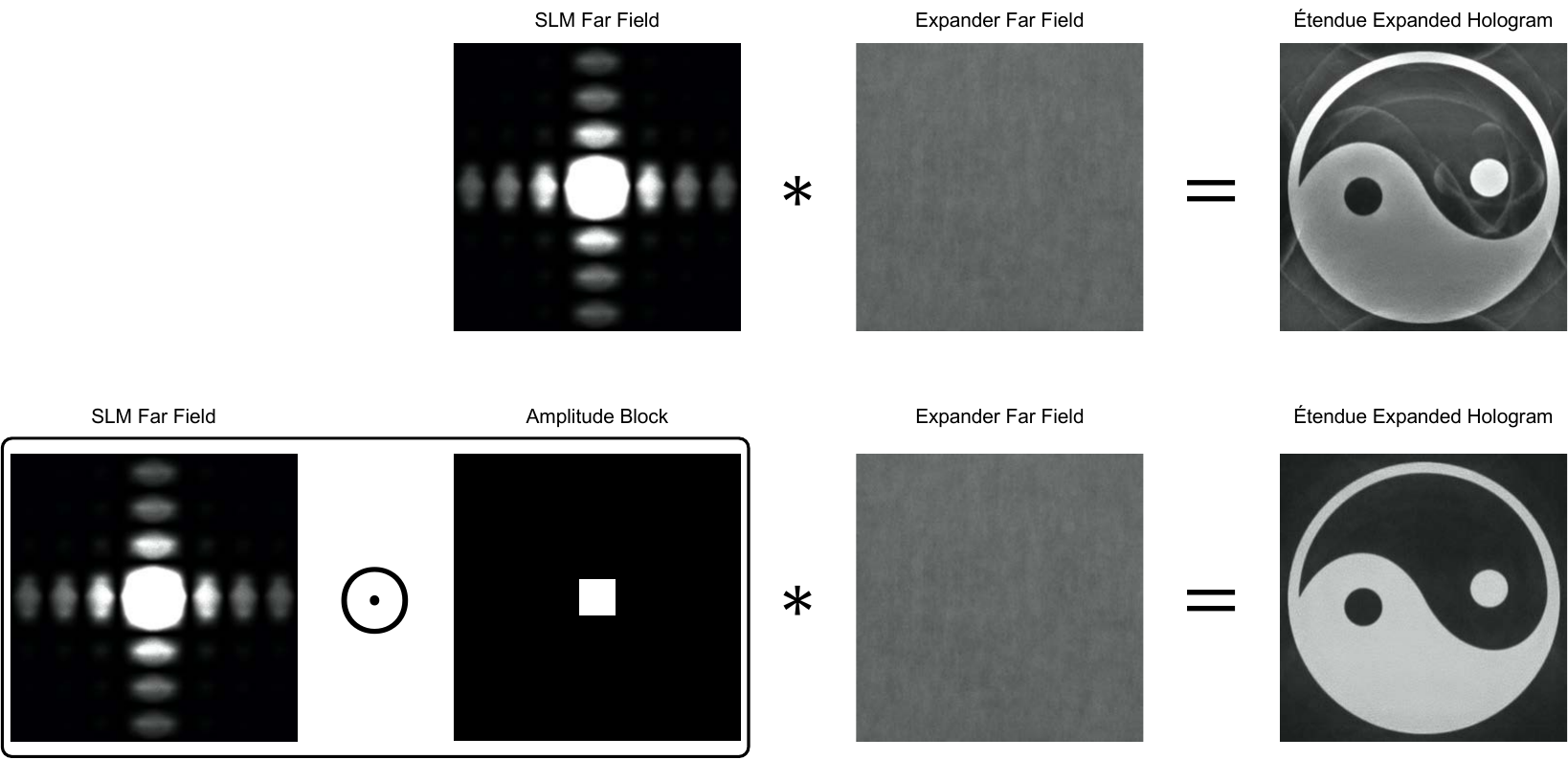}
    \fi
	\caption{\label{fig:sim_echo}
	Removing undesired copies within \'{e}tendue expanded holograms. \'{E}tendue expanded holograms contain undesirable copies because the SLM's far field wavefront contains higher order echoes. Placing a square amplitude block within the 4F system will remove the SLM's higher order echoes. This in turn removes the undesirable copies within the \'{e}tendue expanded hologram.
}
\end{figure*}

\begin{figure*}
	\centering
    	\if\loadFigures0
    \else
    \includegraphics[width=\linewidth]{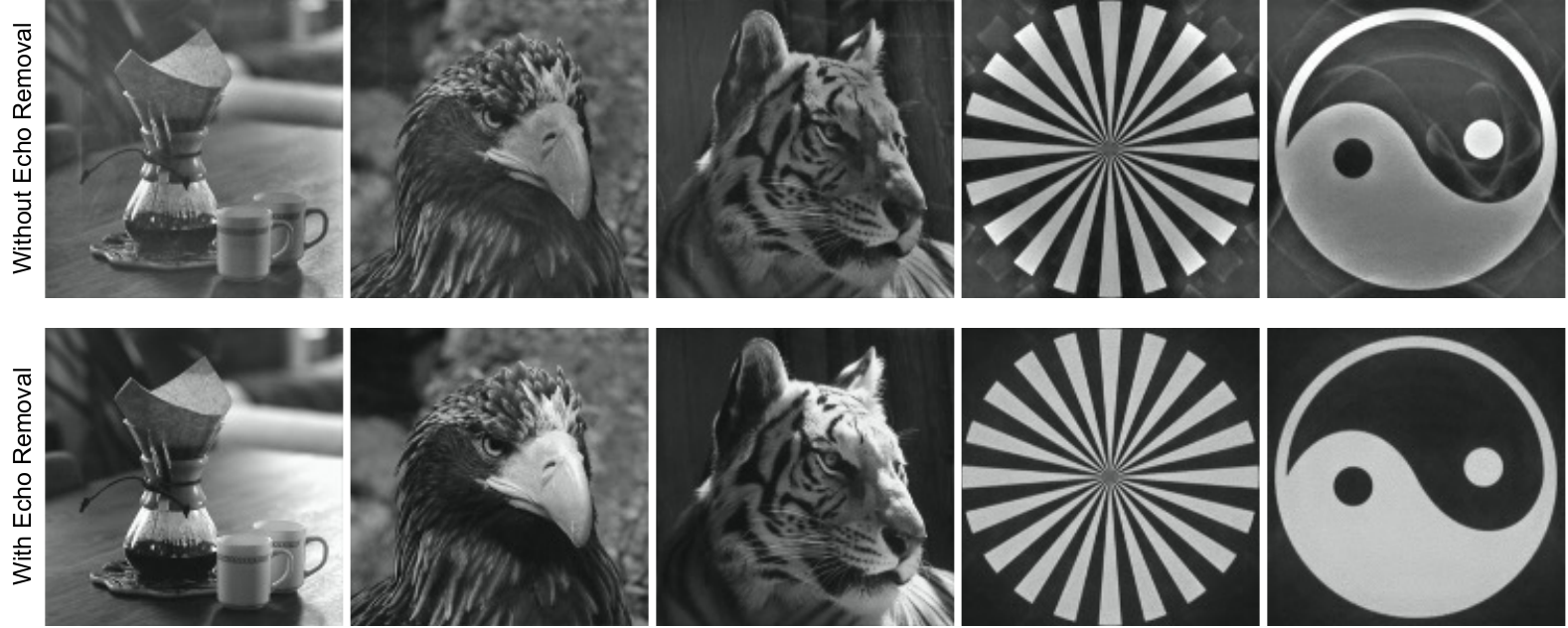}
    \fi
	\caption{\label{fig:sim_echo_results}
	Simulated results for \'{e}tendue expanded holograms with and without higher order echo filtering, see text for detailed description.
}		
\end{figure*}

\clearpage

\section*{Supplementary Note 8: Expander Fabrication}
\label{sec:microscope}
The expanders are physically realized as diffractive optical elements (DOE). The laser writer used for etching the stamp is capable of producing 8 levels of quantization. We set the heights of the DOE to have full $2\pi$ phase range modulation at $\SI{660}{nm}$. Therefore, the 8 height levels of the DOE correspond to $\SI{0}{um}$, $\SI{0.1624}{um}$, $\SI{0.3247}{um}$, $\SI{0.4871}{um}$, $\SI{0.6495}{um}$, $\SI{0.8118}{um}$, $\SI{0.9742}{um}$, and $\SI{1.1366}{um}$. Note that this gives us greater than $2\pi$ phase range for the other wavelengths.

The total mask area of the DOEs spans $\SI{8.192}{mm}\times\SI{8.192}{mm}$. This same area stays constant across all \'{e}tendue expansion factors. For $16\times$ \'{e}tendue expansion the size of each pixel on the DOE is $\SI{8}{um}\times\SI{8}{um}$, corresponding to $1024\times1024$ total pixels. For $64\times$ \'{e}tendue expansion the size of each pixel on the DOE is $\SI{4}{um}\times\SI{4}{um}$, corresponding to $2048\times2048$ total pixels.

\begin{figure*}
	\centering
    	\if\loadFigures0
    \else
    \includegraphics[width=0.9\linewidth]{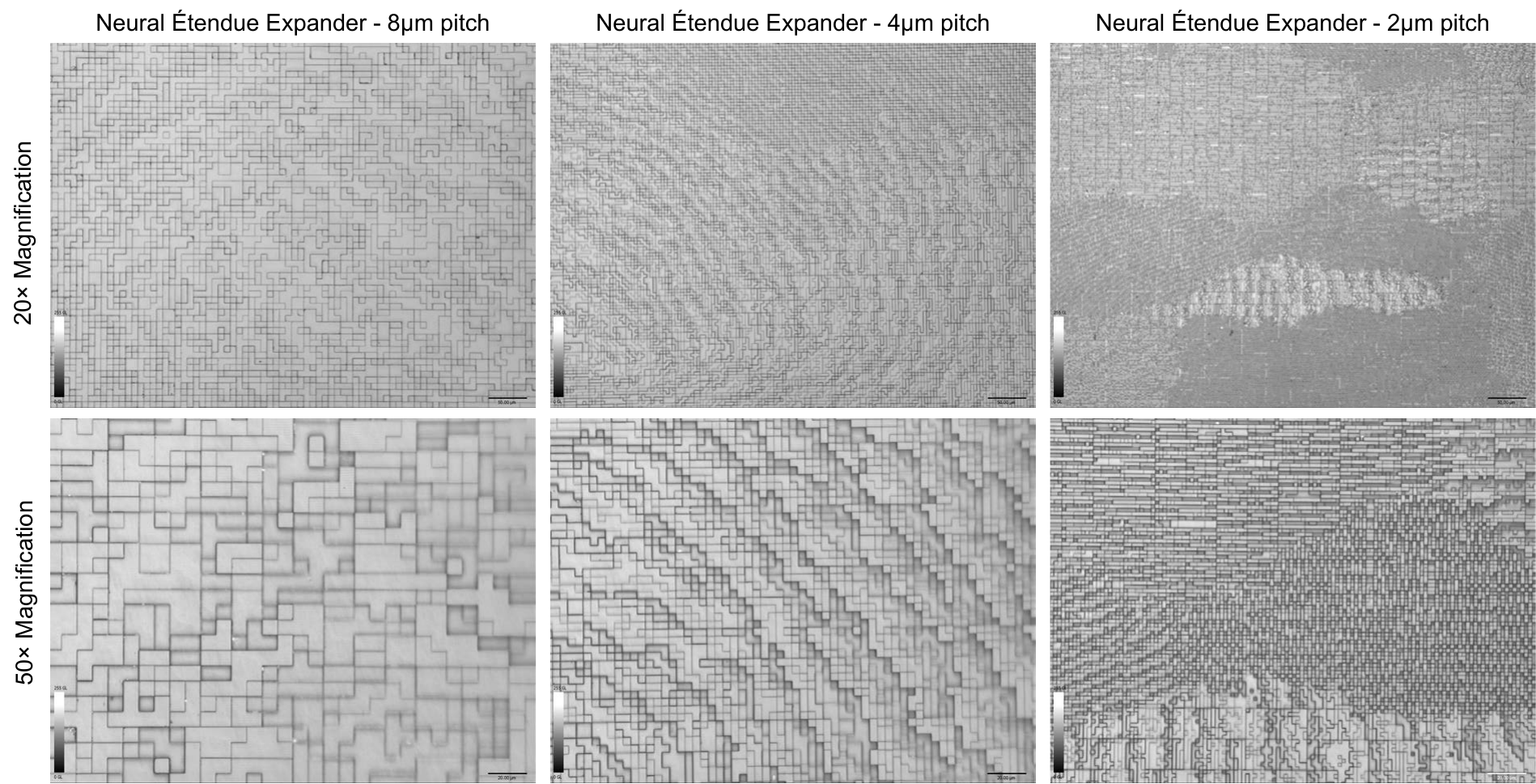}
    \fi
	\caption{\label{fig:fab_nee}
	Confocal microscope imaging of neural \'{e}tendue expander DOEs. These images are recorded at $20\times$ and $50\times$ magnification. Note that these DOEs fully utilize the 8 level quantization which results in several different height levels in these DOEs. Scale bars are shown on the bottom right.
}		
\end{figure*}

\begin{figure*}
	\centering
    	\if\loadFigures0
    \else
    \includegraphics[width=0.9\linewidth]{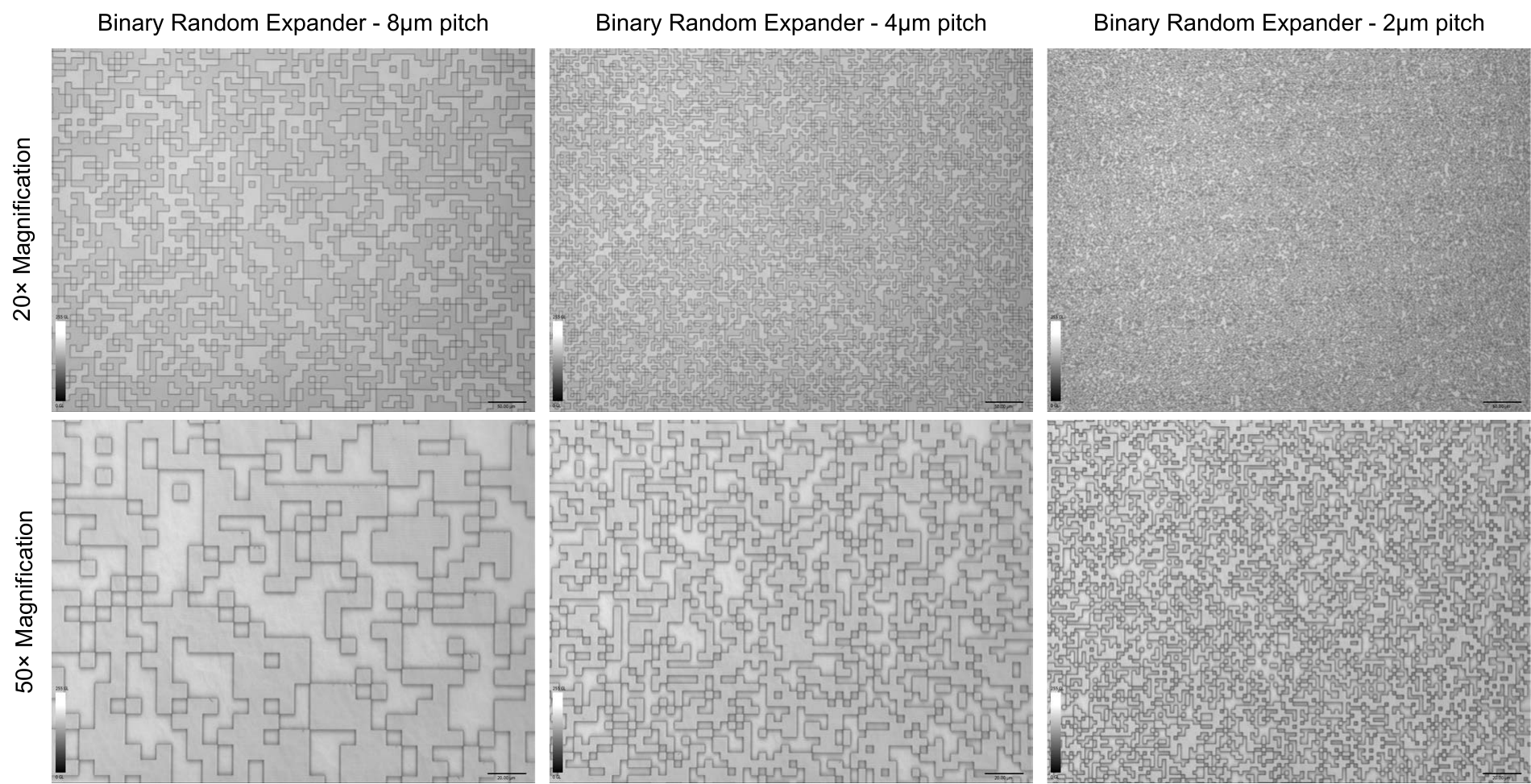}
    \fi
	\caption{\label{fig:fab_random}
	Confocal microscope imaging of random expander DOEs. These images are recorded at $20\times$ and $50\times$ magnification. Note that these DOEs only consist of 2 levels, $\SI{0}{um}$ and $\SI{0.6495}{um}$. Scale bars are shown on the bottom right.
}		
\end{figure*}

\begin{figure*}
	\centering
    	\if\loadFigures0
    \else
    \includegraphics[width=0.9\linewidth]{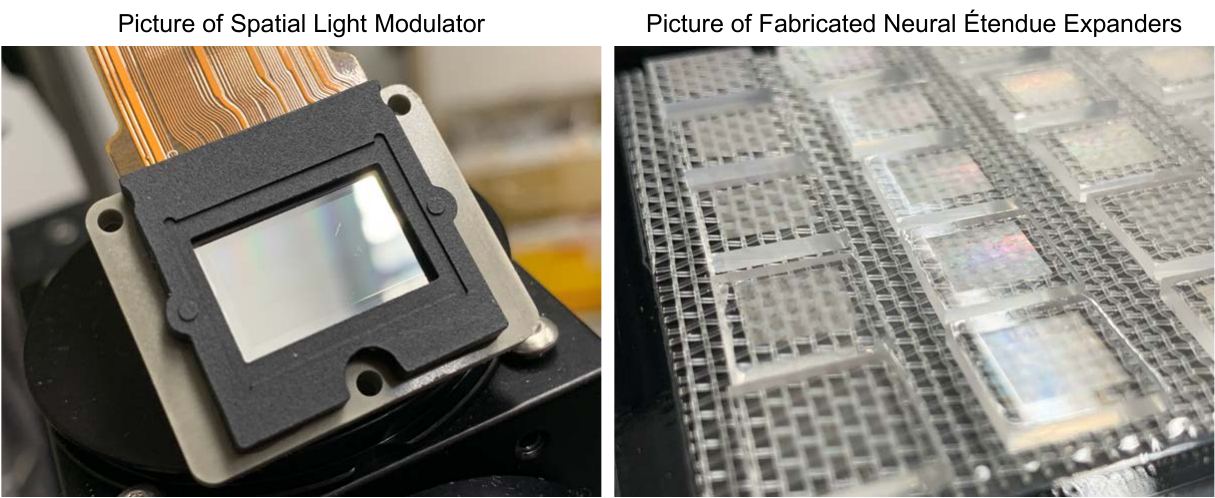}
    \fi
	\caption{\label{fig:fab_picture}
	Photographs of the fabricated DOEs. This picture shows the stamped resin on top of the glass substrate. A picture of the SLM is shown for comparison.
}		
\end{figure*}

Microscope images of the fabricated DOEs corresponding to the optimized neural \'{e}tendue expanders are shown in Fig.~\ref{fig:fab_nee}. Microscope images of the fabricated DOEs corresponding to the random patterns are shown in Fig.~\ref{fig:fab_random}. Pictures of the DOEs are shown in Fig.~\ref{fig:fab_picture}. These images were captured on a Leica DCM 3D confocal microscope.

\clearpage

\clearpage

\section*{Supplementary Note 9: Experimental Setup}
\label{sec:hardware}
We built an experimental setup to validate the neural \'{e}tendue expanders. The setup consists of two main components: the SLM and the expander. 

The holograms are physically realized by illuminating the SLM with a coherent, collimated laser. A single Fourier transforming lens is placed after the expander in order to produce the Fourier holograms. Finally, we use a camera to take pictures of the holograms. A schematic of this setup is shown in Supplementary Figure~\ref{fig:hardware_picture} and a picture of the physical prototype is shown in Supplementary Figure~\ref{fig:hardware_schematic}. A full list of parts is shown in Supplementary Table~\ref{tab:equipment}. Note that our DC blocks are custom made by Frontrange Photomasks, a picture of the DC blocks are shown in Supplementary Figure~\ref{fig:DC_block_picture}.

The current prototype uses a 4F relay system to relay the SLM onto the expander which results in a bulky form factor. In the future a smaller form factor could be achieved by directly integrating the expander onto the SLM. 

\begin{table}
\footnotesize
\caption{Equipment list for building the experimental prototype.}
\label{tab:equipment}
\setlength{\tabcolsep}{3.1pt}
\renewcommand{\arraystretch}{1.0}
\renewcommand\cellalign{ll}
\begin{center}
  \begin{tabular}{llllllllll}
  \toprule
  Equipment                     & Product Name    & Notes \\  \midrule
  Red Laser Diode               & Thorlabs LP685-SF15       & \\\\
  Green Laser Diode             & Thorlabs LP520-SF15       & \\\\
  Blue Laser Diode              & Thorlabs LP450-SF15       & \\\\
  Laser Collimation Lens        & Thorlabs AC254-200-A-ML & Used to collimate the laser source \\\\
  Half Waveplate                & Thorlabs CRM1-LTM       & Used to adjust the polarization of the laser \\\\
  Linear Polarizer              & Thorlabs CRM1   & Used to adjust the polarization of the laser \\\\
  Beam Splitter                 & Thorlabs CCM1-4ER   & \\\\
  Spatial Light Modulator (SLM) & HOLOEYE PLUTO-2 SLM & \\\\
  SLM Mount                     & Thorlabs PY005/M & Used to adjust the orientation of the SLM \\\\
  4F System Lenses              & Pentax SMC FA $\SI{75}{mm}$ $f/2.8$ & Two of the same lenses are required \\\\
  4F System Lens Mounts         & Thorlabs PY005/M & Used to adjust the orientation of the lenses \\\\
  DC Block                      & $\SI{100}{um}$ chrome dot on glass & Used to filter out the SLM's DC term \\\\
  DC Block Holder               & Thorlabs KM100S             & Used to filter out the SLM's DC term \\\\ 
  5-axis translation stage      & Thorlabs Nanomax 300        & Used to adjust the position of the expander \\\\
  Expander Holder               & Thorlabs KM100C        & Provides a 6th-axis for position adjustment \\\\
  Fourier Transforming Lens     & Thorlabs AC508-080-A-ML               & Used to produce the Fourier hologram \\\\
  Imaging Lens                  & Thorlabs AC508-100-A-ML               & \\\\
  Camera Lens                   & Computar C-Mount $\SI{25}{mm}$ Lens   & \\\\
  Camera                        & FLIR Blackfly S BFS-U3-200S6M                & \\\\
  Shear Interferometer          & Thorlabs SI254         & Used to build the system \\\\
  Mirror                        & Thorlabs PF10-03-P01         & Used to build the system \\\\
  Pinhole                       & Thorlabs LMR1AP           & Used to build the system \\ \bottomrule
  \end{tabular}
\end{center}
\end{table}

\begin{figure*}
	\centering
    	\if\loadFigures0
    \else
    \includegraphics[width=\linewidth]{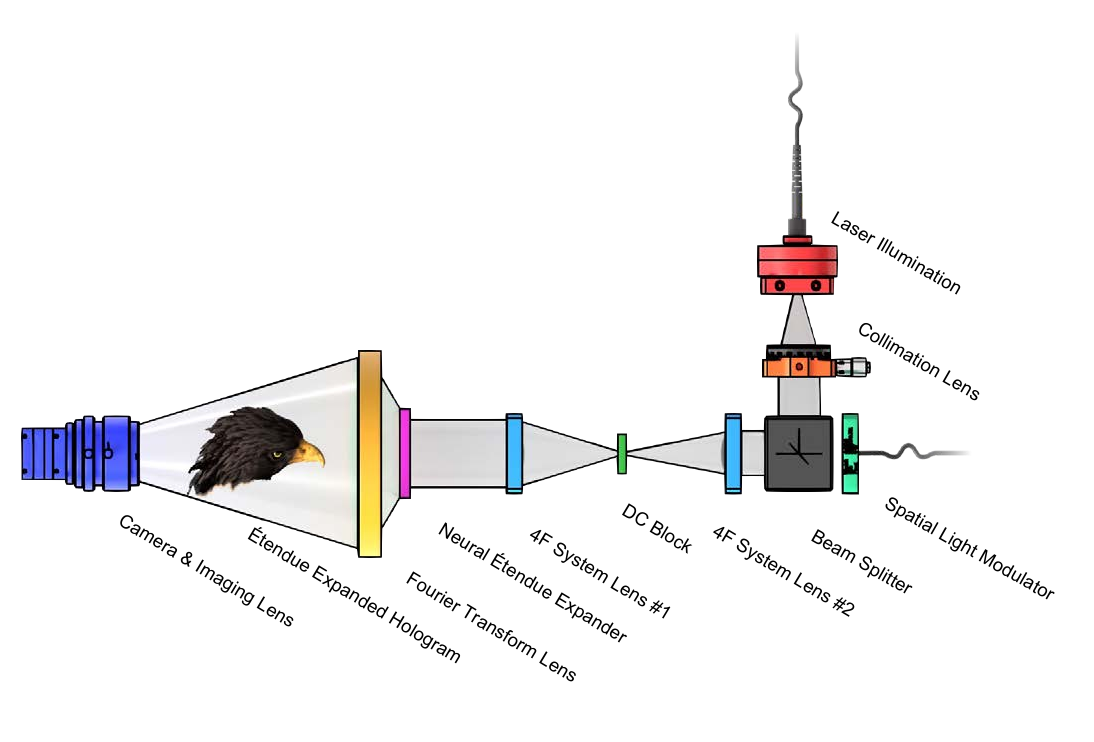}
    \fi
	\caption{\label{fig:hardware_schematic}
	Schematic of experimental setup. Starting from the laser source, the collimating lens turns the laser source into a coherent, collimated beam. We use a half waveplate and a linear polarizer to adjust the polarization of the laser so that it matches the polarization of the SLM. The laser enters the beam splitter cube and the SLM reflects the laser light into the 4F system. The 4F system serves to filter out the SLM DC term via the DC block and also to relay the SLM onto the expander. Once the laser light passes through the expander it is then focused by the Fourier transforming lens to produce the Fourier hologram. This hologram is then imaged by the camera.
}		
\end{figure*}

\begin{figure*}
	\centering
    	\if\loadFigures0
    \else
    \includegraphics[width=\linewidth]{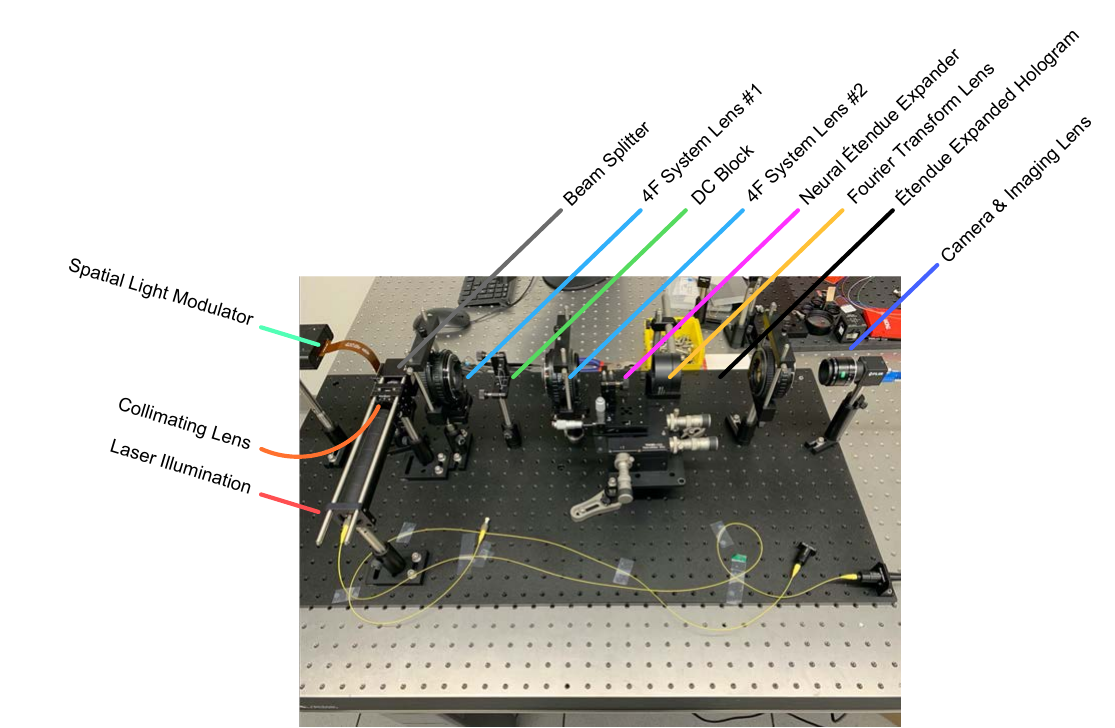}
    \fi
	\caption{\label{fig:hardware_picture}
	Picture of the physical prototype. We constructed this prototype using the parts described in Supplementary Table~\ref{tab:equipment}. The physical realization of each component described in Supplementary Figure~\ref{fig:hardware_schematic} is labeled. The neural \'{e}tendue expander is mounted within the translation mount pointed to by the label. The \'{e}tendue expanded hologram is formed in the space after the Fourier Transforming lens. The hologram cannot be seen in this picture as it can only be seen from the position of the Camera \& Imaging Lens.
}		
\end{figure*}

\begin{figure*}
	\centering
    	\if\loadFigures0
    \else
    \includegraphics[width=0.5\linewidth]{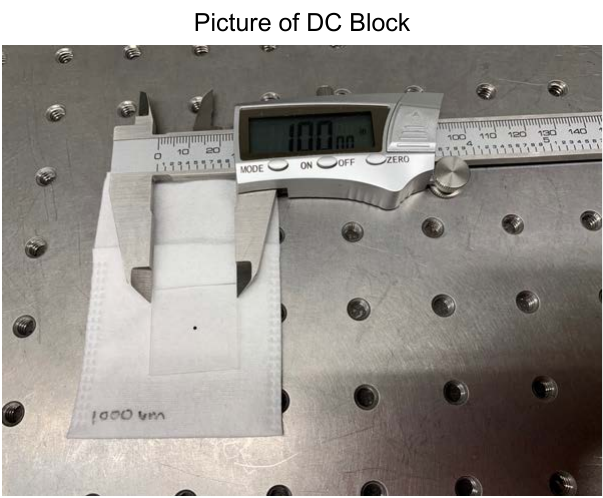}
    \fi
	\caption{\label{fig:DC_block_picture}
	Picture of a DC block. We fill the center of a square piece of glass with a chrome filling. The chrome blocks out the undiffracted DC component of the SLM when placed at the Fourier plane of the 4F system. The diameter of the chrome filling shown in this picture is $\SI{1000}{um}$. For the hardware prototype we used a DC block which has a chrome filling with a smaller diameter of $\SI{100}{um}$.
}		
\end{figure*}

\clearpage

\section*{Supplementary Note 10: Hardware Prototype Construction}
\label{sec:alignment}
In this section, we detail the calibration and alignment procedures we use for the proposed setup.

\paragraph{SLM Calibration} The voltage levels on the reflective SLM need to be calibrated for maximum performance. This is done via the HOLOEYE PLUTO-2 Configuration Manager. Within the configuration manager we set the voltage look-up table to maximize the optical power in the first order at the red wavelength ($\SI{660}{nm}$). The optical power was measured using a power meter (Thorlabs PM100). We also set the low level pixel voltage to $\SI{0.692}{V}$ and the high level pixel voltage to $\SI{1.286}{V}$ to produce a $[0,2\pi]$ phase range at $\SI{660}{nm}$. Setting this phase range for the red wavelength allows for $>2\pi$ phase range for the green and blue wavelengths.

\paragraph{4F System Alignment} We observed that the image quality significantly improves with better alignment. Hence, the following steps ensure that each element in the system is tightly bound and aligned to the main optical axis. Refer also to the equipment list in Table~\ref{tab:equipment}. 

Set up the laser diode and the collimation lens. Use the shear interferometer to validate the collimation of the laser. Place the first 4F system lens after the beam splitter and the first pinhole. Ensure that the lens is mounted so that the optical axis is maintained through the other pinholes. This is done by looking at the laser dots formed on the other pinholes. If the alignment is off then a dot will be formed to the side of the pinhole. It is easiest to see these dots by affixing paper to the pinhole and working with the room lights off. Furthermore, it is also essential to inspect the dots formed via back reflections. Each lens is composed of several internal optical elements. If the alignment is off then internal reflections will occur within those elements and cause reflections back onto the first pinhole. Repeat for all other optical elements in the system. Note that this procedure must be conducted without the DC block in place.

\paragraph{Polarization Calibration} Both the SLM and the laser diode are polarized. The maximum performance is achieved when the polarization states of the two elements are aligned. We will use a half-wave plate (HWP) to turn the polarization of the laser diode. We will use a linear polarizer (LP) after the HWP to eliminate stray polarization states. The goal of this stage is to align the HWP and the LP to the polarization state of the SLM. We used the following steps to align their polarization states.

\begin{figure*}
	\centering
    	\if\loadFigures0
    \else
    \includegraphics[width=\linewidth]{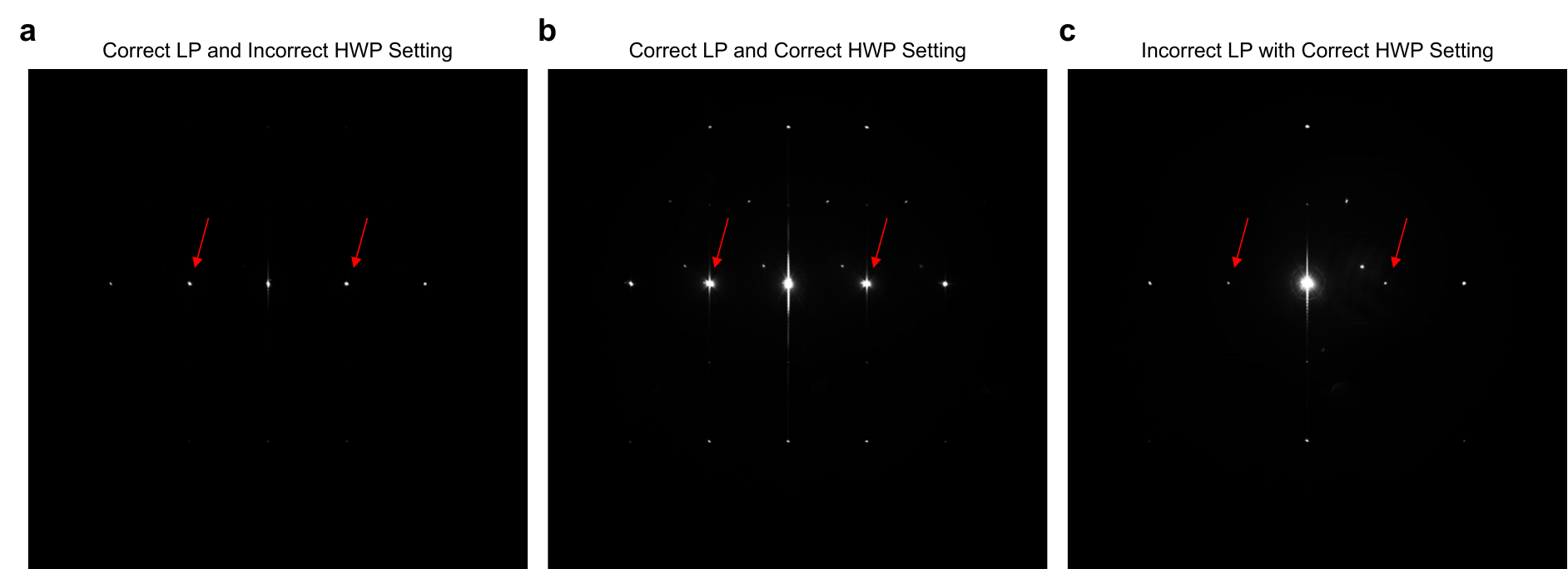}
    \fi
	\caption{\label{fig:polarization}
	\textbf{a} Diffraction pattern when LP is set correctly while the HWP is set incorrectly. \textbf{b} Diffraction pattern when LP and HWP are set correctly. \textbf{c} Diffraction pattern when the LP is set incorrectly while the HWP is set correctly. Observe the size and intensity of the dots indicated by the red arrows.
}		
\end{figure*}

First display the highest frequency blazed grating on the SLM. This pattern is designed by alternately setting each column of the SLM to be $0$ or $\pi$. This pattern will produce a diffraction dot to the right of the main dot produced by the SLM's DC term. An echo of the diffraction dot also will appear to the left of the main dot. We will use the right and left diffraction dots to determine the correct polarization. We turn the LP until the area of the diffracted dots is maximized. The size of the dot should change as you turn the LP. Ignore the setting of the HWP for now. No matter how the HWP is set, changing the LP will always change the size of the diffraction dot. Once the area of the diffracted dots is maximized, we now know that the LP corresponds to the SLM's polarization state, see Figure~\ref{fig:polarization}a. Now we turn the HWP until the intensity of the entire diffraction pattern is maximized, see Figure~\ref{fig:polarization}b. After doing this we know that the HWP and LP must be aligned to the SLM.

As one additional note, turning the LP by $90\deg$ from the correct setting without changing the HWP will create the pattern shown in Figure~\ref{fig:polarization}c. Observe how the area of the diffracted dots changes from Figure~\ref{fig:polarization}b but the main DC dot is mostly unaffected.

\clearpage

\clearpage

\paragraph{Expander Alignment Procedure} After the 4F system is constructed the expander needs to be placed at the precise location of the virtual SLM. This alignment step can be challenging because slight shifts in the expander position will result in a dramatic loss of contrast, oftentimes to the point where there is no discernible signal. Hence, the following steps are used to align the expander.

We construct a 6-axis adjustable translation and rotation stage by combining the Thorlabs KM100C with the Thorlabs Nanomax 300. This stage allows for micrometer-scale movement over the $x$,$y$,$z$ axes and the roll, pitch, yaw axes. We mount the expander by placing it into the Thorlabs KM100C. We adjust the position of the expander until the illumination of the virtual SLM lies within the expander's boundaries.

\begin{figure*}
	\centering
    	\if\loadFigures0
    \else
    \includegraphics[width=\linewidth]{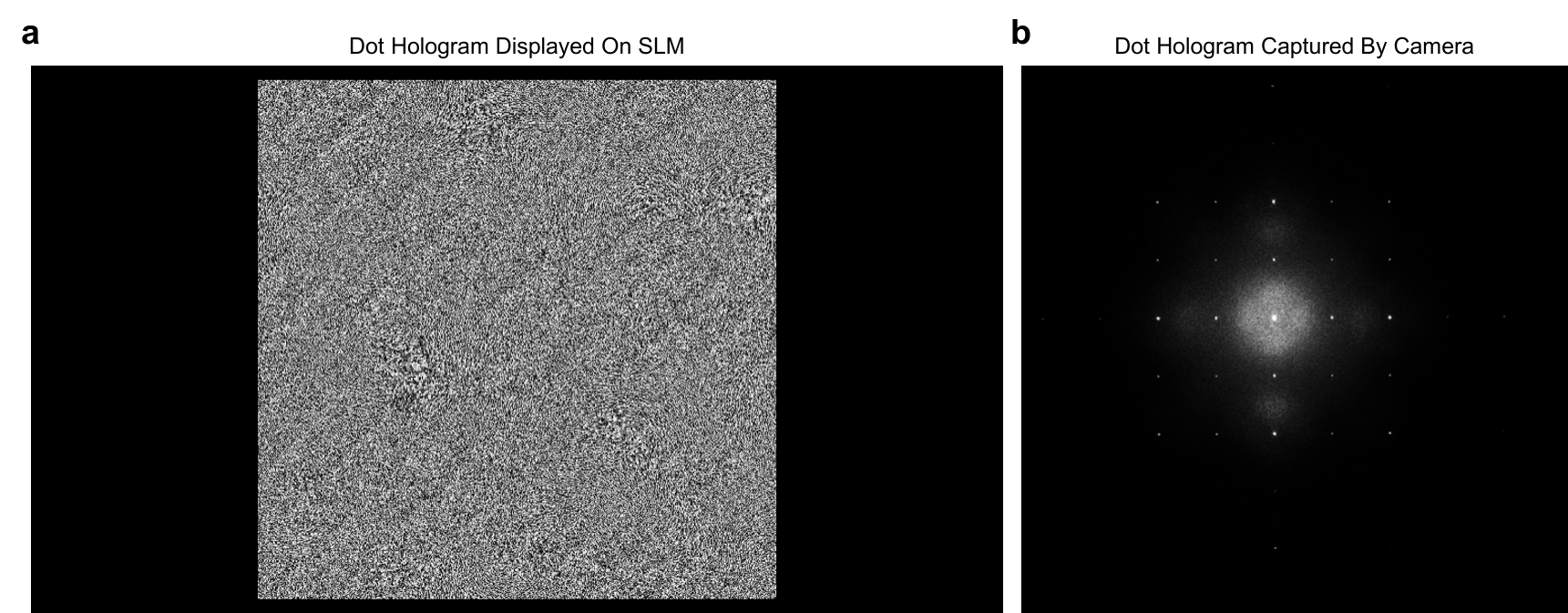}
    \fi
	\caption{\label{fig:dot_pattern}
	\textbf{a} SLM pattern that generates a single dot hologram produced through CGH. Display this pattern on the SLM when performing alignment. \textbf{b} Example of a dot hologram captured by the camera. Adjust the position of the expander until this dot hologram is shown. Note that the initial contrast of the dot may be low, further refinement of the alignment position is necessary to obtain higher contrast.
}		
\end{figure*}

\begin{figure*}
	\centering
    	\if\loadFigures0
    \else
    \includegraphics[width=\linewidth]{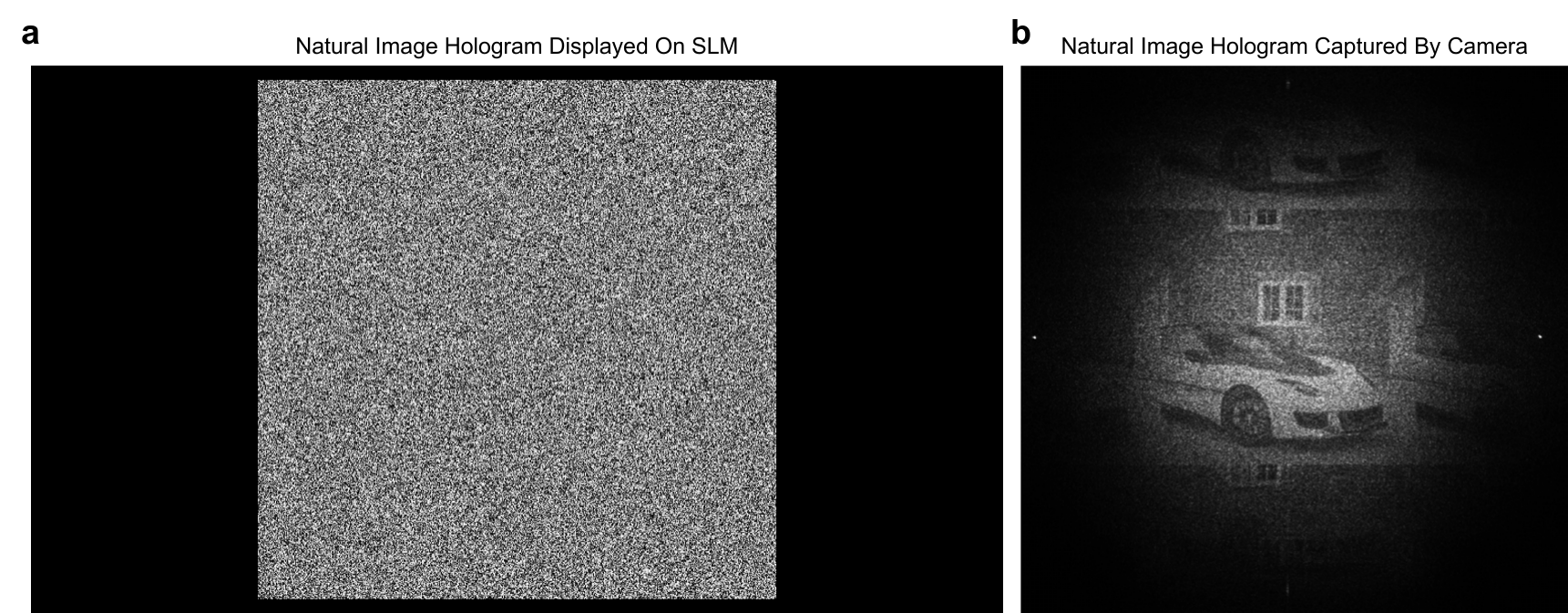}
    \fi
	\caption{\label{fig:natural_image_pattern}
	\textbf{a} SLM pattern that generates a natural image hologram produced through CGH. Display this pattern on the SLM when performing alignment. \textbf{b} Example of a natural image hologram captured by the camera. Note that the initial contrast may be low, further refinement of the alignment position is necessary to obtain higher contrast.
}		
\end{figure*}

We solve for an SLM pattern that will generate a hologram that contains a single dot. Do not use any binning on the SLM. We adjust the stage knobs until the dot is seen. Refer to Figure~\ref{fig:dot_pattern} for a capture of the dot. We then solve for an SLM pattern that will generate a hologram corresponding to a natural image. We display that SLM pattern on the SLM and we align the stage until the contrast of the hologram is maximized. Refer to Figure~\ref{fig:natural_image_pattern} for an example.

The reason that we begin with a dot hologram instead of going directly to a natural scene hologram is because the dot hologram is the most tolerant to alignment error. Therefore, even if the alignment is a little off the dot will be visible, whereas for natural scenes that will not be the case.

Note that the position of the relayed virtual SLM after the 4F system is slightly different for different wavelengths. Thus, the expander needs to be adjusted slightly for different colors for this version of the hardware prototype. Building a system that removes this wavelength variance is a next step.

\clearpage

\clearpage
\bibliographystyle{naturemag}
\bibliography{reference}